\documentclass[sigconf,screen]{acmart}

\usepackage{graphicx}
\usepackage{textcomp}
\usepackage{xcolor}
\usepackage{xspace}
\usepackage{url}
\usepackage{bm}
\usepackage{tabularx}
\usepackage{color}
\usepackage{colortbl}
\usepackage[switch]{lineno}
\usepackage{subcaption}
\usepackage{caption}
\usepackage{tcolorbox}
\usepackage{bbding}
\usepackage{multicol}
\usepackage{amsmath}
\usepackage{hyperref} 
\usepackage{url} 
\usepackage{float} 

\usepackage{braket}
\usepackage{multicol}

\definecolor{light-gray}{gray}{0.8}
\definecolor{light-red}{RGB}{255,155,155}
\definecolor{light-blue}{RGB}{0,155,255}
\definecolor{codegreen}{rgb}{0,0.6,0}
\definecolor{codegray}{rgb}{0.5,0.5,0.5}
\definecolor{codepurple}{rgb}{0.58,0,0.82}
\definecolor{backcolour}{rgb}{0.95,0.95,0.92}

\newcommand{\tool}[1]{\textsc{#1}\xspace}
\newcommand{\bench}[1]{#1\xspace}
\newcommand{\mytool}{\tool{AutoCodeRover}}
\newcommand{\acr}{\tool{ACR}}
\newcommand{\swe}{\bench{SWE-bench}}
\newcommand{\swelite}{\bench{SWE-bench lite}}
\newcommand{\swedevin}{\bench{SWE-bench Devin subset}}
\newcommand{\sweagent}{\tool{Swe-agent}}

\newcommand{\devin}{\tool{Devin}}

\definecolor{sblue}{rgb}{0.36, 0.54, 0.66}

\definecolor{fig4green}{RGB}{151, 208, 119}
\definecolor{fig4blue}{RGB}{169, 196, 235}
\newcommand{\link}{\textcolor{sblue}{\url{https://github.com/nus-apr/auto-code-rover}}\xspace}

\usepackage[colorinlistoftodos,prependcaption,textsize=tiny]{todonotes}

\usepackage{enumerate}
\usepackage[shortlabels]{enumitem}

\usepackage{tcolorbox}

\usepackage{multirow}
\usepackage{pbox}
\usepackage{tabularx}

\usepackage{listings}
\newcommand{\lstbg}[3][0pt]{{\fboxsep#1\colorbox{#2}{\strut #3}}}
\lstdefinestyle{mystyle}
{
    language = Python,
    basicstyle = {\ttfamily \color{main-color}},
    keywordstyle = {\color{blue}},
    keywordstyle = [2]{\color{blue}},
    keywordstyle = [3]{\color{yellow}},
    keywordstyle = [4]{\color{teal}},
    morekeywords = [3]{<<, >>},
    morekeywords = [4]{++},
    basicstyle=\ttfamily\footnotesize,
    commentstyle=\color{gray}\ttfamily,
    morecomment=[f][\lstbg{red!20}]-,
    morecomment=[f][\lstbg{green!20}]+,
    morecomment=[f][\lstbg{yellow!20}]++,
    morecomment=[f][\lstbg{yellow!20}]--,
    morecomment=[f][\textit]{@@},
    texcl=false
}

\lstdefinestyle{prompt_style}
{
    language = {},
    keywordstyle = {\color{blue}},
    keywordstyle = [2]{\color{blue}},
    keywordstyle = [3]{\color{yellow}},
    keywordstyle = [4]{\color{teal}},
    morekeywords = [3]{<<, >>},
    morekeywords = [4]{++},
    basicstyle=\ttfamily\footnotesize,
    commentstyle=\color{gray}\ttfamily,
    morecomment=[f][\lstbg{red!20}]-,
    morecomment=[f][\lstbg{green!20}]+,
    morecomment=[f][\lstbg{yellow!20}]++,
    morecomment=[f][\lstbg{yellow!20}]--,
    morecomment=[f][\textit]{@@},
    texcl=false,
    numbers=none,
    breakindent=0pt
}

\usepackage{etoolbox,environ}
\newtoggle{isArxivVersion}

\definecolor{fpbackcolor}{RGB}{242,242,242}
\definecolor{diffrem}{RGB}{202, 45, 49}
\definecolor{diffincl}{RGB}{0, 135, 90}
\definecolor{codepink}{RGB}{237, 2, 140}

\newboolean{showcomments}
\setboolean{showcomments}{true} 
\ifthenelse{\boolean{showcomments}}{
  \newcommand{\nbc}[3]{
    {\textcolor{#3}{\small{\bfseries{#1:\ }}\textit{#2}}}}
}{
  \newcommand{\nbc}[3]{}
}

\usepackage{amsfonts}

\usepackage{braket}

\usepackage{pifont}

\usepackage{xcolor}
\usepackage{soul}

\newcommand{\snapshot}[1]{12th July 2024\xspace}

\newcommand{\revised}[1]{{{#1}\xspace}}

\setcopyright{rightsretained}
\acmDOI{10.1145/3650212.3680384}
\acmYear{2024}
\copyrightyear{2024}
\acmISBN{979-8-4007-0612-7/24/09}
\acmConference[ISSTA '24]{Proceedings of the 33rd ACM SIGSOFT International Symposium on Software Testing and Analysis}{September 16--20, 2024}{Vienna, Austria}
\acmBooktitle{Proceedings of the 33rd ACM SIGSOFT International Symposium on Software Testing and Analysis (ISSTA '24), September 16--20, 2024, Vienna, Austria}
\acmSubmissionID{issta24main-p1370-p}
\received{2024-04-12}
\received[accepted]{2024-07-03}

\begin{document}

\title{AutoCodeRover: Autonomous Program Improvement}

\author{Yuntong Zhang}
\orcid{0009-0005-1664-7110}
\affiliation{%
  \institution{National University of Singapore}
  \country{}
}
\email{yuntong@comp.nus.edu.sg}

\author{Haifeng Ruan}
\orcid{0009-0008-1080-4770}
\affiliation{%
  \institution{National University of Singapore}
  \country{}
}
\email{hruan@comp.nus.edu.sg}

\author{Zhiyu Fan}
\orcid{0000-0002-8165-9493}
\affiliation{%
  \institution{National University of Singapore}
  \country{}
}
\email{zhiyufan@comp.nus.edu.sg}

\author{Abhik Roychoudhury}
\orcid{0000-0002-7127-1137}
\affiliation{%
  \institution{National University of Singapore}
  \country{}
}
\email{abhik@comp.nus.edu.sg}






\begin{abstract}
Researchers have made significant progress in automating the software development process in the past decades. Automated techniques for issue summarization, bug reproduction, fault localization, and program repair have been built to ease the workload of developers. Recent progress in Large Language Models (LLMs) has significantly impacted the development process, where developers can use LLM-based programming assistants to achieve automated coding. Nevertheless, software engineering involves  the process of program improvement apart from coding, specifically to enable 
software maintenance (e.g. program repair to fix bugs) and 
software evolution (e.g. feature additions). In this paper, we propose an automated approach for solving Github issues to autonomously achieve program improvement. In our approach called \mytool, LLMs are combined with sophisticated code search capabilities, ultimately leading to a program modification or patch. In contrast to recent LLM agent approaches from AI researchers and practitioners, our outlook is more software engineering oriented. We work on a program representation (abstract syntax tree) as opposed to viewing a software project as a mere collection of files.
Our code search exploits the program structure in the form of classes/methods to enhance LLM's understanding of the issue's root cause, and effectively retrieve a context via iterative search. The use of spectrum-based fault localization using tests, further sharpens the context, as long as a test-suite is available. \revised{Experiments on the recently proposed SWE-bench-lite (300 real-life Github issues) show increased efficacy in solving Github issues (19\% on SWE-bench-lite), which is higher than the efficacy of the recently reported \sweagent. Interestingly, our approach resolved 57 GitHub issues in about 4 minutes each (pass@1), whereas developers spent more than 2.68 days on average. In addition, \mytool achieved this efficacy with significantly lower cost (on average, \$0.43 USD), compared to other baselines.} 
We posit that our workflow enables autonomous software engineering, where, in future, auto-generated code from LLMs can be autonomously improved.
\end{abstract}






\maketitle

\section{Beyond Automatic Programming}

Automating software engineering tasks has long been a vision among software engineering researchers and practitioners. One of the key challenges has been the handling of ambiguous natural language requirements, in the process of automatic programming. In addition, there has been progress in some other software engineering activities such as automated test generation \cite{bohme2020fuzzing,cadar-sen}, automated program repair \cite{cacm19}, and so on.

Recent progress in large language models (LLMs) and the appearance of tools like Github Copilot \cite{copilot} hold significant promise in automatic programming. 
This progress immediately raises the question of whether such automatically generated code can be trusted to be integrated into software projects, and if not, what improvements to the technology are needed. One possibility is to automatically repair generated code to achieve {\em trust}. This brings out the importance of automating program repair tasks towards achieving the vision of autonomous software engineering. 


Given this motivation of automating program repair, and the large number of hours developers often spend manually fixing bugs, we looked into the possibility of fully autonomous program improvement. Specifically, we feel that bug fixing and feature addition are the two key categories of tasks that a development team may focus on when maintaining an existing software project. To achieve this goal, we proposed an approach that augments LLM with context knowledge from the code repository. We call our tool \mytool.

Technically our solution works as follows. Given a real-life GitHub issue, LLM first analyzes the attached natural language description to extract keywords that may represent files/classes/methods/code snippets in the codebase. Once these keywords are identified, we employ a stratified strategy for the LLM agent to retrieve code context by invoking multiple necessary code search APIs at one time with the keyword combinations as arguments (e.g., search\_method\_in\_file). These code search APIs are running locally based on AST analysis and are responsible for retrieving code context such as class signatures and method implementation details from a particular location in the codebase. By collecting the project context with code search APIs, LLM refines its understanding of the issue based on the currently available context. Note that our code context retrieval will proceed in an iterative fashion. The LLM agent directs the navigation and decides which code search APIs to use (i.e. where/what to retrieve code) in each iteration based on the current available context returned from the previous API calls.
\mytool then enquires whether there is sufficient project context, and subsequently uses the collected context to derive the buggy locations. The patch construction is then handled by another LLM agent which considers the buggy locations as well as all the context collected so far for those locations. 

\mytool also can leverage  debugging techniques such as spectrum-based fault localization (SBFL)~\cite{sbfl-survey} to decide more precise code search APIs for context retrieval if a test suite accompanying the project is available. SBFL primarily considers the control flow of the passing and failing tests and assigns a suspiciousness score to the different methods of the program. The LLM agent may prioritize retrieving context from particular methods and classes if the fault localization result is provided, e.g., when a method appears both in the issue description and in the output of fault localization.
In the last step, \mytool may perform patch validation using available tests, to determine whether the patch produced by \mytool passes the tests in the given test-suite. Otherwise, \mytool will rerun the patch generation with a retry limit until a correct patch is found.

\paragraph{Contributions}
Our contribution lies in the effective use of code search to make software engineering processes like program repair autonomous. \revised{Instead of viewing the codebase as a collection of files, we posit \mytool as an approach that gleans specification from software structure, which is used to guide the patching. The code search in \mytool is a vehicle for inferring specification from program structure.} We demonstrate this capability by autonomously solving GitHub issues.
We report favorable experimental results on the \swelite dataset \cite{jimenez2024swebench}. We achieve 
19\% efficacy on the \swelite with 300 GitHub issues.
Overall, we see the need to endow a {\em software engineering oriented outlook} to the recent flurry of activity on LLM agents for software engineering, which mostly has only an AI flavor. The angle of software engineering can be conveyed by the following five dimensions.
\begin{itemize}[leftmargin=1em]\itemsep0em 
\item We work on program representations (abstract syntax tree or AST) as opposed to viewing a software project as a collection of files. We posit that working on program representations like AST will be useful in autonomous software engineering workflows. 
\item To solve GitHub issues, we 
focus on code search in a way that resembles the activity of a human software engineer. So we try to use the program structure - classes, methods, code snippets - in searching for relevant code context. 
This leads to a more effective usage of the context provided to LLM. 
\item We posit that higher efficacy of automated repair is more important than time efficiency, {\em as long as the time is within a threshold}. It is well-known in empirical software engineering research that time limits of 30-60 minutes for automated repair are tolerable, based on extensive developer surveys in the field \cite{icse22}. We thus report 19\% efficacy on \swelite in solving GitHub issues, within 4 minutes. We can compare this time limit of 4 minutes to the average time of 2.68 days to fix the GitHub issues manually.

\item One should be able to exploit debugging techniques like test-based fault localization to guide the search for code in resolving GitHub issues. We show that the use of fault localization in setting the code context leads to an increase in the efficacy of \mytool in solving GitHub issues.

\item Finally, it is also useful to examine how many of the solved GitHub issues are producing acceptable patches. We study the patches produced by \mytool and report that 2/3 of the autonomously produced patches from \mytool are correct and acceptable. We note that this aspect has not been reported by Devin~\cite{devin} and SWE-agent~\cite{yang2024sweagent}.
\end{itemize}


\section{Relevant Literature}

\begin{figure*}[!t]
  \centering
  \includegraphics[width=0.9\textwidth]{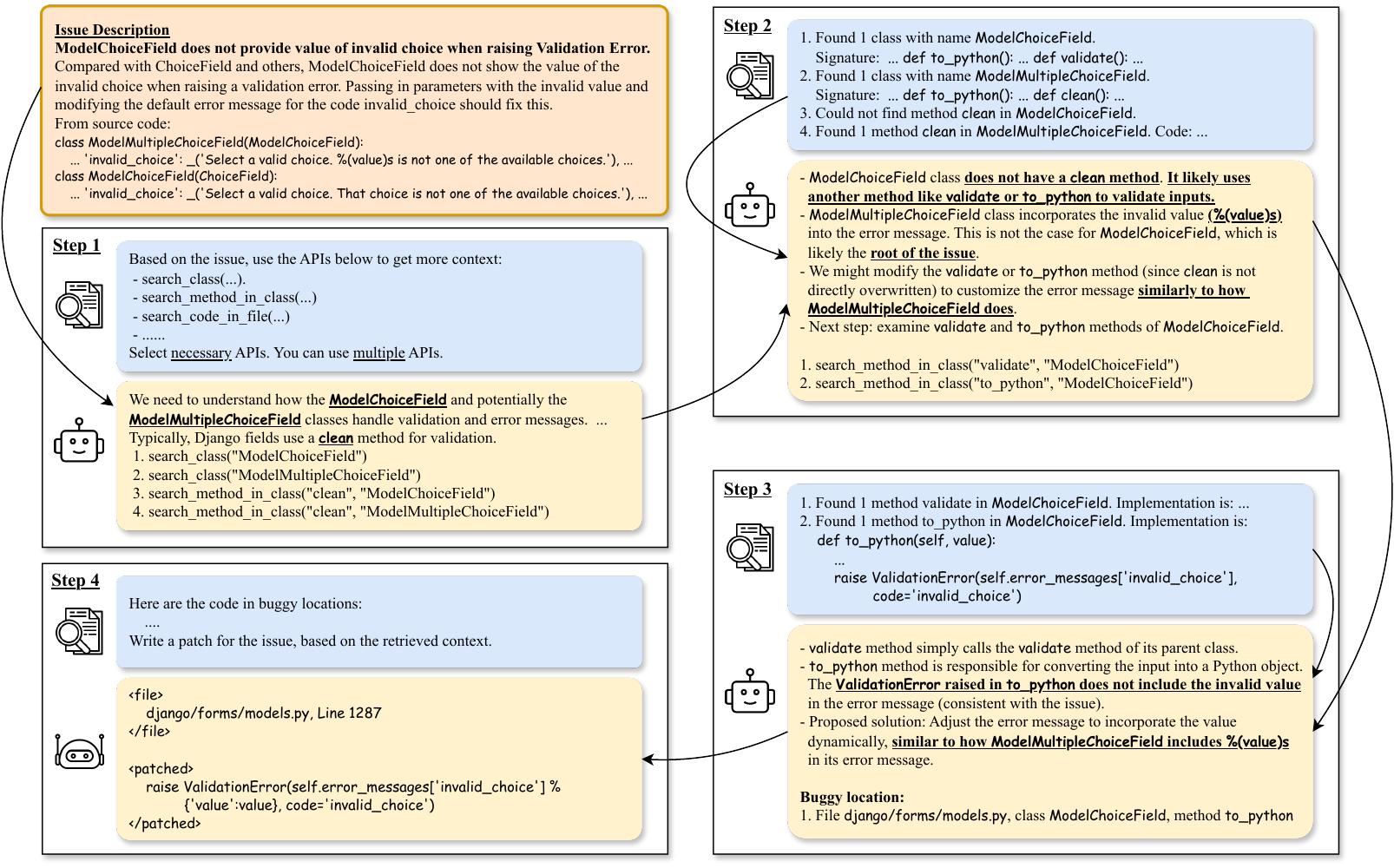}
  \caption{Issue description of django-13933 and \mytool's workflow on it.}
  \label{fig:motivating-example}
\end{figure*}

\subsection{Program Repair}

Test-suite based automated program repair (APR) has attracted significant attention in the last decade~\cite{cacm19}. These techniques aim to generate a patch for a buggy program to pass a given test-suite. APR techniques typically include search-based, semantic-based, and pattern/learning-based APR. Search-based APR techniques like GenProg~\cite{genprog} take a buggy program and generate patches using predefined code mutation operators, or search for a patch over the patch space that passes the given test suite.  Semantics-based APR techniques~\cite{semfix,angelix} generate patches by formulating a repair constraint that needs to be satisfied based on a given test suite specification, and then solving the repair constraint to generate patches. Learning-based APR techniques~\cite{lutellier2020coconut, cure, recoder} often train a deep learning model with large code repositories and are guided by a specific representation of code syntax and semantics to predict the next tokens that are most likely to be correct patch. There are also works~\cite{koyuncu2019ifixr,tan2024crossfix,motwani2023better} that tried to leverage GitHub issues and bug reports to improve APR effectiveness.


 Recent work~\cite{fan2023automated,jiang2023impact,xia2023automated} have shown the use of LLMs for automated program repair. This line of work often assumes the buggy program statements are given (i.e. perfect fault localization assumption) and focuses on constructing APR-specific prompts that guide LLM to generate a patch for the selected buggy program statements multiple times until a patch that passes all tests is found. The typical work targets repairing functional~\cite{fan2023automated, prenner2022can, wang2023rap}, type~\cite{chow2024pyty} bugs, and even software vulnerabilities~\cite{fu2022vulrepair,jin2023inferfix,pearce2023examining} in different languages with open-source/commercialised models (e.g., CodeT5, ChatGPT, GPT-4, etc). However, obtaining buggy locations for a large project is an essential and challenging task in resolving real-life bug reports. 


APR techniques have been successfully deployed in industries for domain-specific bug fixing~\cite{williams2023user,marginean2019sapfix,bader2019getafix}.
However, a long-standing challenge for the APR techniques is to resolve general real-life software issues from scratch. The above APR techniques rely on a high-quality test suite which is not always available in the real world and they do not leverage the valuable natural language specification from the original problem description. To address these challenges and achieve autonomous software engineering, we focus on resolving GitHub issues from a real-life dataset.

\subsection{LLM Agents for SE and Dataset}
\label{sec:swe}
\swelite~\cite{jimenez2024swebench} is a benchmark that aims to evaluate the capabilities of large language models in resolving end-to-end real-life software engineering tasks. The benchmark consists of 300 real-life software engineering task instances collected from the repositories of 11 popular large Python projects~\footnote{https://www.swebench.com/index.html} (e.g., \href{https://github.com/django/django}{django}, \href{https://github.com/sympy/sympy}{sympy}). Each \swelite task instance contains a pair of Github issue and corresponding pull requests. The GitHub issue either reports a bug to be fixed or requests to implement new features. The pull request includes the code changes made by human developers to resolve the issue and test cases that prevent the issue. Unlike traditional code generation tasks in HumanEval~\cite{codex} and MBPP~\cite{alphacode} benchmark, resolving a \swelite instance is particularly challenging because it requires automatically generating code changes that address the problem in a GitHub issue for a matured large code repository based only on the issue description. More specifically, the process may involve a series of complex tasks like reasoning the target bug location across files in the code repository, analyzing the root cause of the issue, proposing bug-fixing strategies, and eventually writing a patch that passes all the test cases added in the pull request. There are a few primary attempts at tackling tasks in \swelite. \devin~\cite{devin} is named the first AI software engineer that can solve various software engineering tasks, including building a project from scratch, bug-fixing/feature-addition for existing projects. However, it is a close-sourced commercial tool, and its details are not available. \sweagent~\cite{yang2024sweagent} is a concurrent work against \mytool. It designed an agent-computer interface (ACI) that allows LLM agents to execute basic file operations via shell commands to achieve interaction between the LLM engine and a software repository. Compared to \sweagent, we posit  \mytool as an approach that gleans specification from software structure, which is used to guide the patching. Instead of viewing the codebase as a collection of files.



\section{Motivating Example}
\label{sec:motivating}

\begin{figure*}[!t]
  \centering
  \includegraphics[width=\textwidth]{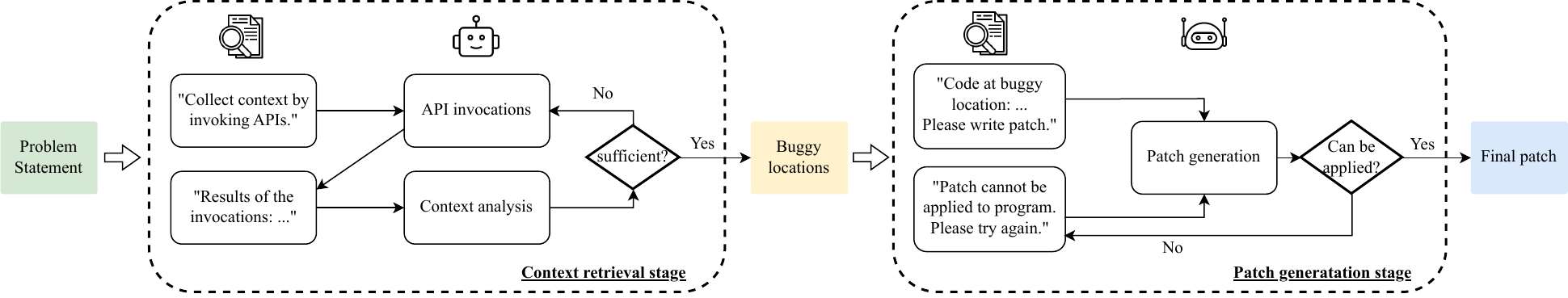}
  \caption{Overall workflow of \mytool.}
  \label{fig:overall-workflow}
\end{figure*}

\begin{figure}[h]
  
\begin{lstlisting}[numbers=none]
""" AutoCodeRover patch. """
 except (ValueError, TypeError, self.queryset.model.DoesNotExist):
-    raise ValidationError(self.error_messages['invalid_choice'],
-        code='invalid_choice')
+    % Include the invalid value in the error message
+    raise ValidationError(
+        self.error_messages['invalid_choice'] % {'value': value},
+        code='invalid_choice')
""" Developer patch. """
 except (ValueError, TypeError, self.queryset.model.DoesNotExist):
-    raise ValidationError(self.error_messages['invalid_choice'], 
-        code='invalid_choice')
+    raise ValidationError(
+        self.error_messages['invalid_choice'],
+        code='invalid_choice',
+        params={'value': value},)
\end{lstlisting}
  \caption{\mytool and developer patch for Django-13933.}
  \vspace{-10pt}
  \label{fig:example-patch}
\end{figure}


In this section, we illustrate how  our tool \mytool can collect code context and generate a patch from an issue description and the corresponding project code-base. We show an example of a feature addition task.
Figure~\ref{fig:motivating-example} demonstrates the workflow of \mytool on an issue submitted to the Django issue tracker\protect\footnotemark. 
This issue is classified as ``New feature'' in the issue tracker, and is included in \swelite with the id ``django-13933''.
The first part of Figure~\ref{fig:motivating-example} shows the issue description (the code part is simplified for brevity).
This issue requests adding support to the \texttt{ModelChoiceField} class, so that it ``shows the value of the invalid choice when raising a validation error''.

\footnotetext{\url{https://code.djangoproject.com/ticket/32347}}

With this issue description, \mytool operates in two stages - \emph{context retrieval} (Step 1-3 in Figure~\ref{fig:motivating-example}) and \emph{patch generation} (Step 4).
First, a context retrieval LLM agent is instructed to collect the relevant code context related to this issue, from a local copy of the Django project codebase.
This retrieval is achieved by inferring relevant names (e.g. \texttt{ModelChoiceField}) and searching for them in the Abstract Syntax Tree (AST) of the project.
A set of retrieval APIs are provided to the agent.
In Step 1, the agent identifies the classes \texttt{ModelChoiceField} and \texttt{ModelMultipleChoiceField} might be relevant based on the issue description, and infers that a \texttt{clean} method might be also of interest for validation-related issues.
It then invokes the \texttt{search\_class} and \texttt{search\_method\_in\_class} APIs to retrieve more information about them.


In Step 2, API invocations return the signature of the classes and implementation of the methods found.
Results of invocation \#3 reveals the absence of method \texttt{clean} in the class \texttt{ModelChoiceField}, which helps to \emph{refine} the agent's understanding from Step 1.
The returned class signature from invocation \#1 also reveals two more relevant methods \texttt{to\_python} and \texttt{validate}, which were not mentioned in the issue description.
This suggests the retrieval should be performed \emph{iteratively} in multiple steps, so that results from a previous search can become arguments of the following search.
In this example, the agent then iteratively invokes \texttt{search\_method\_in\_class} on the two newly revealed methods.
Furthermore, by referencing results from multiple invocations, the agent can infer that \texttt{ModelMultipleChoiceField} incorporates the invalid value into the message with \texttt{\%(value)s}, and methods in \texttt{ModelChoiceField} can be modified \emph{similarly} to \texttt{ModelMultipleChoiceField}.

In Step 3, the agent receives the implementation of \texttt{validate} and \texttt{to\_python} methods.
Among the two methods, it selects \texttt{to\_python} to be the more suitable place to make changes, since \texttt{to\_python} raises the relevent exception and does not include the invalid value.
At this point, the retrieval agent deems the collected code context as sufficient for understanding the issue and drafting the patch.
The identified buggy location, together with the gathered context and analysis so far, is passed to another patch generation agent.
This agent is instructed to write patches following the format described in Step 4 (see yellow box) of Figure~\ref{fig:overall-workflow}. 
In Step 4, a patch is written to allow a \texttt{value} to be integrated into the error message, by utilizing \%-formatting in Python.
Figure~\ref{fig:example-patch} shows the patch generated by \mytool and the developer.
Although written in a different way compared to the developer patch, this patch achieves similar effect and passes the developer provided test-suite for this issue.




\section{AI Program Improvement Framework}
\label{sec:methodology}

In this section, we discuss the design of \mytool. 
\mytool is a system incorporating AI agents for program improvement tasks in large software projects. 
\mytool is designed to work in a realistic software development lifecycle, in which users submit issue reports to a software repository describing a bug, and the project maintainers craft a patch to resolve the issue.
With a submitted issue, \mytool autonomously analyzes the submitted issue, retrieves the relevant code context in the software project, and generates a patch. 
This patch can then be vetted by human developers. If such a tool can automatically handle a certain percentage of the issues awaiting developers' attention, manual efforts are reduced.
\begin{table*}[t]
    \centering
    \caption{List of Context Retrieval APIs.}
    \label{tab:api}
    \begin{tabular}{>{\raggedright}p{4.2cm}|>{\raggedright\arraybackslash}p{5.4cm}|>{\raggedright\arraybackslash}p{6cm}}
        \hline
        \textbf{API name} & \textbf{Description} & \textbf{Output} \\
        \hline
        search\_class (cls) & Search for class \texttt{cls} in the codebase. & Signature of the searched class. \\
        \hline
        search\_class\_in\_file (cls, f) & Search for class \texttt{cls} in file \texttt{f}. & Signature of the searched class. \\
        \hline
        search\_method (m) & Search for method \texttt{m} in the codebase. & Implementation of the searched method. \\
        \hline
        search\_method\_in\_class (m, cls) & Search for method \texttt{m} in class \texttt{cls}.  & Implementation of the searched method. \\
        \hline
        search\_method\_in\_file (m, f) & Search for method \texttt{m} in file \texttt{f}. & Implementation of the searched method. \\
        \hline
        search\_code (c) & Search for code snippet \texttt{c} in the codebase. & +/- 3 lines of the searched snippet \texttt{c}. \\
        \hline
        search\_code\_in\_file (c, f) & Search for code snippet \texttt{c} in file \texttt{f}. &  +/- 3 lines of the searched snippet \texttt{c}. \\
        \hline
    \end{tabular}
\end{table*}

\subsection{Overview}
\label{overview}
We first describe the overall stages \mytool operates in, and will proceed in more details in the rest of this section.
The overall workflow of \mytool is shown in Figure~\ref{fig:overall-workflow}. \mytool takes in as input a problem statement $P$ of the issue to be resolved, and a codebase $C$ of the corresponding software project.
This problem statement $P$ contains the title and description of the issue, as shown in Section~\ref{sec:motivating}.  
From a problem statement written in natural language, \mytool analyzes the requirement from it and proceeds in two main stages, which are \emph{context retrieval} and \emph{patch generation}. 

In the \emph{context retrieval} stage, \mytool employs an  LLM agent to navigate through a potentially large codebase $C$ and extract the relevant code snippets relevant to $P$. 
This navigation is facilitated by a set of \emph{context retrieval APIs} (Section~\ref{api}), which enables an LLM to retrieve information about the project (e.g. class signatures) and actual code snippets (e.g. method implementations).
The context retrieval agent directs this navigation, and decides which retrieval APIs to use based on the current available context.
In order to make the LLM-directed navigation more ``controlled'', we devise a stratified stategy of invoking the retrieval APIs (Section~\ref{layered}).
This stratified strategy instructs the LLM to only invoke necessary retrieval APIs based on the available information, and iteratively changes the set of retrieval APIs used when more code-related context are returned from the previous API calls.

Once the context retrieval agent has gathered sufficient context about the issue, \mytool proceeds to the \emph{patch generation} stage (Section~\ref{patch-generation}). 
In this stage, \mytool employs another LLM agent to extract more precise code snippets from the retrieved context, and craft a patch based on the extracted code snippets.
This patch generation agent is instructed to craft a patch in a specific format, and if the produced patch does not follow the format specification or cannot be applied to the original codebase, the agent enters a retry-loop which terminates after a pre-configured number of attempts.
Finally, \mytool outputs a patch that attempts to resolve the original issue.

We note that the workflow of \mytool discussed so far only requires the problem statement $P$ and codebase $C$ as input, and do not require any program specifications such as testcases. 
However, when testcases are available (e.g. provided by developers or generated from another tool), execution-based information and program analysis techniques can be integrated into the \mytool framework (Section~\ref{analysis}).
For example,  statistical fault localization  \cite{sbfl-survey} tool can be used to reveal more relevant methods beyond those mentioned in the problem statement.
The additionally revealed code context can thereby influence the set of retrieval APIs invoked in the \emph{context retrieval} stage.
Moreover, with testcases available, the patch generation agent can employ an additional patch validation step in its retry-loop when crafting a patch.
In the remainder of this section, we discuss in greater detail the components of \mytool and their design considerations.



\begin{figure}
    \centering
    \includegraphics[width=\columnwidth]{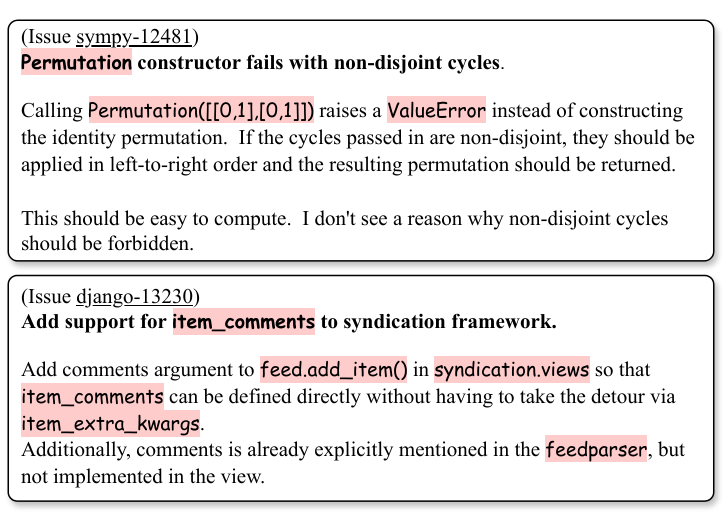}
    \caption{Issue description of sympy-12481 and django-13230. ``Hints'' are highlighted.}
    \label{fig:issue-examples}
\end{figure}

\begin{figure*}[t]
  \centering
  \includegraphics[width=0.95\textwidth]{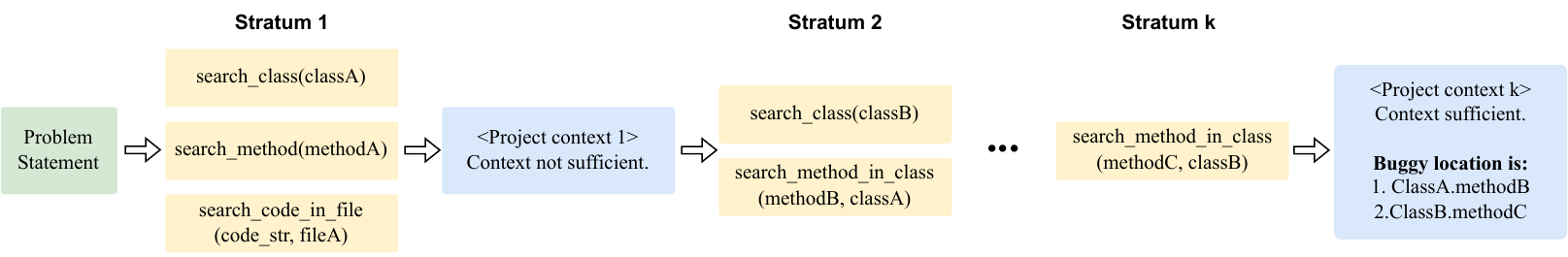} 
  \caption{Stratified search with retrieval APIs for context gathering.}
  \label{fig:stratified}
\end{figure*}

\subsection{Context Retrieval APIs}
\label{api}

In the \emph{context retrieval} stage, the basic components are a set of APIs which an LLM agent can use to gather relevant code context and snippets from the codebase. 
For typical software project issues, we observe that users often mention some ``hints'' on which part of the codebase is relevant. 
These hints can be the names of the relevant methods, classes, or files, and sometimes also contain short code snippets. 
Although these hints may not directly point to the precise location for code modification, they often reveal code context in the project that is relevant to the current issue. Figure~\ref{fig:issue-examples} illustrates two real-world issues submitted to the sympy and django projects, respectively, and the ``hints'' are highlighted. In sympy-12481, the class \texttt{Permutation} is mentioned twice; in django-13230, multiple hints are mentioned, such as code snippet \texttt{feed.add\_item()}, package path \texttt{syndication.views}, method name \texttt{item\_extra\_kwargs} and etc.
Based on the type of project and code related hints, we design a set of APIs for an LLM agent to retrieve code context from these hints.
The current set of APIs in \mytool and their outputs are shown in Table~\ref{tab:api}.
Once invoked by the LLM agent, the retrieval APIs search for classes, methods and code snippets in the code-base, and return the results back to the agent.
To avoid forming very lengthy code context that creates distraction during patch generation, we return only necessary information as API outputs.
For example, since the complete definition of a class can be lengthy in large projects, we only return signature of the class as output for \texttt{search\_class} and \texttt{search\_class\_in\_file}. Returning the signature to shorten cntext, is a better approach than cutting off the context at a certain bound.
Upon receiving the class signature, the agent could then invoke another API to search for the relevant methods / snippets inside the class.

\paragraph{Interfacing with LLM} For the LLM agent to invoke the context retrieval APIs, the description and expected output of them are presented to it as part of prompt. When the agent decides to invoke a set of retrieval APIs, it responds with the list of API call names and the corresponding arguments. These retrieval API requests are processed locally by parsing a local codebase of the project into AST and searching over it. Results of locally executing these APIs are returned to the agent, forming the code context. 




\subsection{Stratified Context Search}
\label{layered}

The set of context retrieval APIs listed in Table~\ref{tab:api} serves as building blocks for searching relevant code context. 
With the context retrieval APIs and the LLM-identified keyword ``hints'' from the problem statement, a set of possible \emph{API invocations} can be derived by using the identified keywords as API parameters.
We discuss a few observations on using these API invocations to gather code context, and propose a \emph{stratified} context retrieval process.




Our first observation is that the context retrieval should not be restricted to a single API invocation.
For example, in the issue django-13230 mentioned in Figure~\ref{fig:issue-examples}, if the retrieval starts from the invocation \texttt{search\_method("add\_item")}, the implementation of method \texttt{add\_item} can be considered by the LLM agent as a sufficient context, since it appears to be relevant to the problem statement.  However, searching from only one method can lead to incomplete context for the agent to reason about root cause of the problem.
On the other hand, if all API invocations are executed at once, a large code context can be retrieved, especially when the problem statement mentions many class and method names. This large code context can be difficult for an LLM to comprehend, or may even exceed its context window.


The second observation is that some of the API invocation results provide more elements to build new possible API invocations, which means the process of invoking retrieval APIs should be iterative.
For example, the result of a \texttt{search\_class} call returns the method signatures within the searched class, and an LLM can iteratively invoke method-related APIs afterwards.

With these two observations, we propose a \emph{stratified} search process for invoking context retrieval APIs, as illustrated in Figure~\ref{fig:stratified}.
From a problem statement, stratified search iteratively invokes retrieval APIs to gather project code context, and finally outputs a list of potentially buggy locations to be fixed.
In each stratum, we prompt the LLM agent to select a set of \emph{necessary} API invocations, based on the current context. In stratum 1, the current context only contains the problem statement; in the following strata, the context contains both the problem statement and the code searched so far.
By allowing LLM to select more than one API invocations and also instructing it to only select the necessary ones, we make the best use of the context, building what we deem an {\em optimal} context.

After the API invocations in a stratum are executed, the newly retrieved code snippets are added to the current context.
The LLM agent is then prompted to analyze whether the current context is sufficient for understanding the issue, thereby deciding whether (a) we continue the iterative search process, or (b) we decide on the buggy locations which will be considered for fixing.




\subsection{Analysis-Augmented Context Retrieval}
\label{analysis}



We also investigate how program  debugging techniques can augment our workflow.
Specifically, we integrate the \emph{Spectrum-based Fault Localization} (SBFL) analysis into \mytool to study the effect of such a test-based dynamic  analysis. We make a test-suite $T$ available to \mytool, in addition to the problem statement $P$ and codebase $C$.

\paragraph{Spectrum-based Fault Localization.} The goal of SBFL is to identify the location of software faults~\cite{flsurvey}.
Given a test-suite $T$ containing both passing and failing tests, SBFL considers control-flow differences in the passing and failing test executions, and assigns a suspiciousness score to different program locations. 
This suspiciousness score can be computed with various metrics such as Tarantula~\cite{tarantula} and Ochiai~\cite{ochiai}. 
Program elements (e.g. statements/basic blocks) with the highest suspiciousness scores are identified as likely fault locations. 
SBFL can be performed at different granularities of program elements, such as statements or basic blocks. 
Since the LLM can process reasonably long code snippets, we use \emph{method-level} SBFL in \mytool.
Given an test-suite, SBFL can be used to directly output a few program locations to be repaired. 
However, the accuracy of SBFL highly relies on the quality of the test-suite~\cite{sbfleval} - 
since the SBFL results are effectively an abstraction of the differential of control flows between passing tests and failing tests.
Therefore, instead of replacing the stratified context retrieval with SBFL, we use the SBFL identified methods to augment the search process.
Before \mytool enters the context retrieval stage, we provide the SBFL-identified methods to the LLM agent as ``results from an external analysis tool that identifies suspicous code''.
The main role of SBFL-identified methods is to reveal more ``hints'' on relevant classes and methods beyond those mentioned in the problem statement.
The LLM agent can then use the context retrieval APIs to examine these methods.
Since the SBFL-identified methods are presented to the agent together with the problem statements, the agent can then cross-reference between these two sources of information.
For example, if one of the SBFL-identified method names is more closly related to the problem statement, the LLM is more likely to invoke the \texttt{search\_method} API on this name.
We will demonstrate this observation in Section~\ref{sec:eval-rq2}.




\subsection{Patch Generation}
\label{patch-generation}

In the \emph{patch generation} stage, \mytool employs a \emph{patch generation agent} to use the collected code context to write a patch for the problem statement.
This agent is given the problem statement, the identified buggy locations/methods, and the history of context retrieval, including the invoked APIs, the API results, as well as the previous analysis on the code context made by the context retrieval agent.

As a first step, the patch generation agent retrieves the precise code snippets at the buggy locations from the codebase.
Based on the precise code snippets and other relevant code context, the agent enters a retry-loop of generating patches.
If a generated patch does not follow the specified patch format or could not be applied syntactically to the original program, the agent is prompted to retry.
We also employs a linter to indentify Python-specific syntax issues such as indentation errors in this retry-loop.
The agent is allowed to retry up to a pre-configured attempt limits (currently set to three), after which the best patch so far is returned as output.



\section{Experiment Setup}
\label{sec:setup}

To evaluate the capabilities of  \mytool in resolving real-life software issues, we answer the following research questions.
\begin{description}[leftmargin=*]\itemsep0em 
    \item[RQ1:] To what extent can \mytool automate software issues like human developers?

    \item[RQ2:] Can existing debugging / analysis techniques assist \\ \mytool?
    
    \item[RQ3:] What are the challenges for \mytool and fully automated program improvement in future?
\end{description}

\paragraph{Benchmark} We evaluate \mytool in recently proposed benchmarks \swe and \swelite~\cite{jimenez2024swebench}, comprising 2294 and 300 real-life GitHub issues, respectively.
The only input is the natural language description in the original GitHub issue and its corresponding buggy codebase. Details of \swe and \swelite appear in Section~\ref{sec:swe}.

\paragraph{Baseline and Evaluation Metric} We selected two LLM-based agent systems \devin~\cite{devin}, \sweagent~\cite{yang2024sweagent}
as baselines and compare their performance against \mytool. We directly compare \mytool with \sweagent's original results as it is publicly available at GitHub\footnote{https://github.com/princeton-nlp/SWE-agent}. In contrast to \sweagent, we do not have access to \devin, so we take the most relevant reported result from their technical report~\cite{devinreport}.
To avoid the natural randomness of LLM, we repeat our experiments three times. We report the result with the \mytool@1 and \mytool@3 annotations (i.e. pass@1, pass@3 metrics~\cite{codex} respectively). 
We use (1) the percentage of resolved instances, (2) average time cost, and (3) average token cost to evaluate the effectiveness of the tools. 
These evaluation metrics represent overall effectiveness, time efficiency, and economic efficacy in resolving real-world GitHub issues.


\paragraph{Implementation and Parameters} We use the state-of-the-art OpenAI GPT-4 (gpt-4-0125-preview) as foundation inference model for \mytool. In \mytool, the GPT-4 model is responsible for selecting search APIs to retrieve codebase context, refining the issue description, and writing a final patch. For parameters of GPT-4, we set a low temperature=0.2, max\_tokens=1024 to produce relatively deterministic results and enable sufficient reasoning length for \mytool, and all other parameters remain as per default. \mytool terminates either a patch is generated or the context retrieval stage repeats ten times. Experiment results and artifacts for \mytool can be found at \link.

\paragraph{System Environment} All experiments are conducted on an x86\_64 Linux server with Ubuntu 20.04 installed. The correctness of generated patches by \mytool is evaluated on the official \swe docker environment.

\section{Evaluation}
\label{sec:evaluation}

\subsection{RQ1: Overall Effectiveness}
\label{sec:eval-rq1}

\begin{figure}[ht]
  \centering
    \includegraphics[width=0.75\columnwidth]{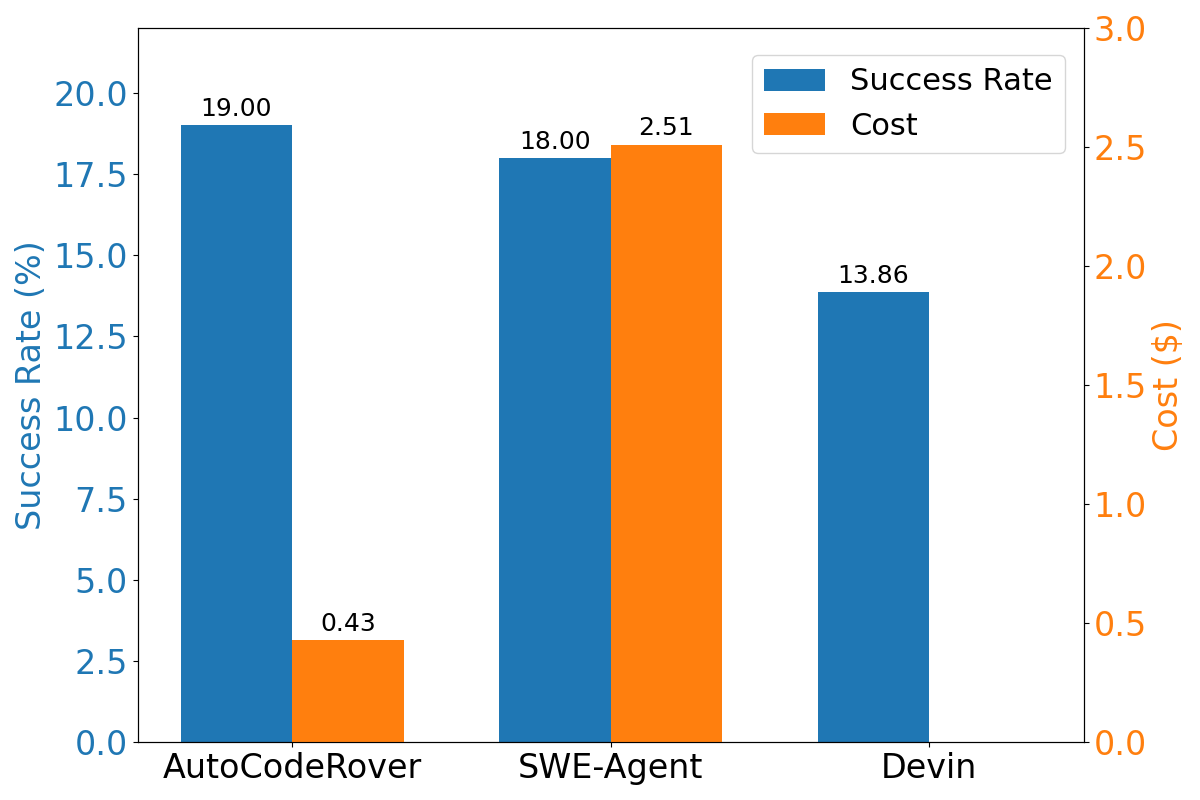}
  \caption{Task-resolving success rates (\%) and average costs per task in USD of \mytool and baselines.}
      \label{fig:success-cost}
\end{figure}

We first measure the overall effectiveness of \mytool and baselines with the number of resolved task instances in \swe. With the goal of understanding to what extent the current AI systems can automatically resolve real-life software issues, the only inputs we provided are the natural language issue description and a local code repository checked out at the erroneous version. 
We repeated the experiment of \mytool three times to avoid randomness and presented 
the results in pass@1 and pass@3 metrics~\cite{codex}, which are denoted as \mytool@1 and \mytool@3 respectively. 
When reporting time and token/cost for \mytool@3, we report the time and cost required for running each task three times. Since \devin was evaluated on a random 25\% subset of \swe~\cite{devin}, we also report results of \mytool on this subset (we refer to this as ``\swedevin'').
Table~\ref{tab:overall-result} shows the overall result of \mytool in \swelite, full \swe and \swe Devin subset and Figure~\ref{fig:success-cost} shows a visual efficacy summary of \mytool's comparison with \sweagent, \devin.




\paragraph{Results on \swe.} \revised{The results reported in Table~\ref{tab:overall-result} indicate that in the full \swe, \mytool@1 resolved 12.42\% task instances with 248 seconds per task, and \mytool@3 resolved 17.96\% task instances taking a time of 701 seconds per task, which is at par with 12.47\% resolved tasks by the concurrent work \sweagent, according to their technical report~\cite{yang2024sweagent}. We also performed another round of experiments with \mytool on \swelite (300 instances). \mytool@1 resolved 19\% task instances in taking an average time of 195 seconds, whereas \sweagent resolved 18\% task instances. 
Moreover, we also investigated \mytool@3 in \swelite,
in which the percentage of resolved task instances increased to 26\%. The results of \mytool@3 in \swe imply that \mytool may be complemented by running multiple times, highlighting the possibility of improving \mytool with an iterated generate-and-validate process in the future, if any program specification is available.}






\paragraph{Comparison with Devin.} To compare with \devin, we report \mytool's result on the 570 task instances (random 25\% subset of \swe) \devin~\cite{devin} was evaluated on. The results of \devin are taken from their technical report~\cite{devinreport}. In this \devin particular subset, \mytool successfully resolved 12.63\% task instances on average in pass@1, and 18.77\% of the task instances in pass@3, which is higher than \devin. Besides, the time taken by \mytool is much smaller than \devin.

\begin{figure}[t]
  \centering
        \includegraphics[width=0.6\columnwidth]{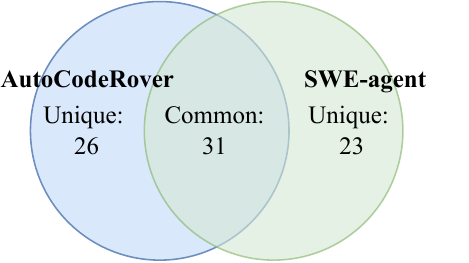}

  \caption{Venn diagrams of resolved tasks instances by \mytool and \sweagent, on \swelite.}
        \label{fig:venn2}
\end{figure}

\paragraph{Detailed Comparison with \sweagent} In the rest of RQ1, we compare \mytool with \sweagent on \swelite using the highlighted \mytool@1 result.
We further analyzed the commonly and uniquely resolved instances between \mytool and \sweagent in Figure~\ref{fig:venn2}, and we found that \mytool and \sweagent complement each other in different scenarios. \mytool uniquely resolved 26 task instances, which benefited from the fine-grained code context search at the AST level to precisely locate the bug locations (e.g., django-13401 searches three necessary methods at one time). On the other hand, the main reason that \mytool failed on the 23 unique instances resolved by \sweagent is unimplemented search APIs (e.g., search\_file invoked in django-12286). When such unimplemented APIs are invoked, \mytool returns an error message to the LLM, but the LLM may still fail to invoke valid search APIs in later attempts.
This implies more robust search APIs are desired for future improvement of \mytool. 


\begin{table}[t!]
    \centering
    \tabcolsep3pt 
    \caption{Overall Result of \mytool and baselines on full \swe, \swedevin, and \swelite. "-" indicates data not available.}
    \label{tab:overall-result}
       \begin{tabular}{l|l|l|l}
        \hline
        Tools & Resolved Tasks & Avg Time & Avg Tokens\\\hline\hline

        \multicolumn{4}{c}{Reported result on \swelite (size=300)}\\\hline
        
        \sweagent~\cite{yang2024sweagent} & 18.00\% (54) & - & 245k (\$2.51)\\
        \mytool@1 & 19.00\% (57) & 195 & \bf 37k (\$0.43) \\
        \mytool@3 & 26.00\% (78) & 520 & 112k (\$1.30) \\
        \acr-sbfl & 22.00\% (66) & 250 & 40k (\$0.47) \\
        \hline\hline
         \multicolumn{4}{c}{Reported result on full \swe (size=2294)}\\\hline
        \sweagent~\cite{yang2024sweagent} &  12.47\% (286) & - & 240k (\$2.46) \\
        \mytool@1 & 12.42\% (285) & 248 & \bf 39k (\$0.45) \\
        \mytool@3 & 17.96\% (422) & 701 & 120k (\$1.39) \\

        \hline\hline
        \multicolumn{4}{c}{Reported result on \swe \devin subset (size=570)}\\\hline
        \sweagent~\cite{yang2024sweagent} & 13.51\% (77) & - & 234k (\$2.40) \\
        \devin~\cite{devin,devinreport}\footnotemark & 13.86\% (79) & > 600 & -\\
        \mytool@1 & 12.63\% (72) & 238 & \bf 37k (\$0.42) \\
        \mytool@3 & 18.77\% (107) & 692 & 117k (\$1.36) 
        \\

        \hline
    \end{tabular}
\end{table}
\footnotetext{The reported result of \devin is evaluated on a random 25\% subset of full \swe}

\begin{figure}[ht]
  \centering
    \includegraphics[width=0.75\columnwidth]{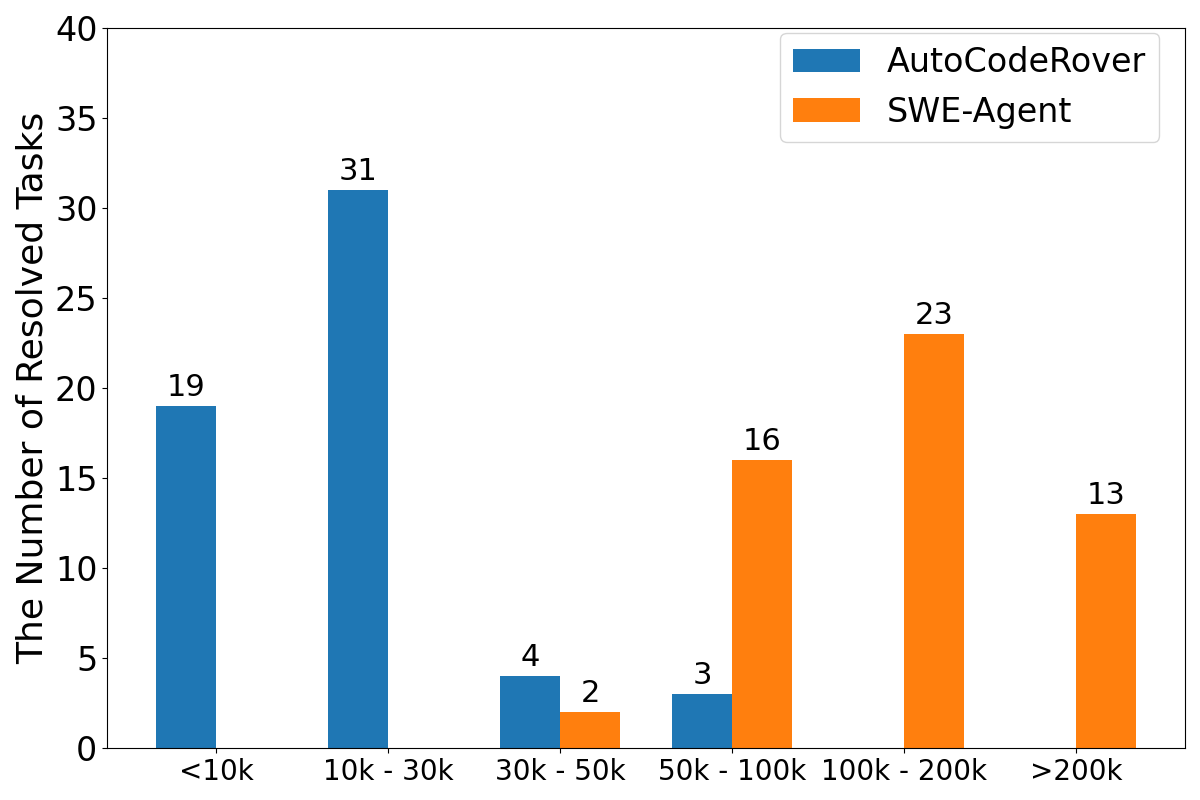}
  \caption{The number of resolved tasks and token cost distributions of \mytool and \sweagent.}
      \label{fig:token-resolve}
\end{figure}

 \paragraph{Time / Token Cost} Today, the price of invoking state-of-the-art LLMs such as GPT-4 and Claude-3-Opus are still very expensive. 
 Hence, we are also interested in assessing the feasibility of deploying \mytool in the real world in terms of time and economic cost. \revised{Figure~\ref{fig:success-cost} shows the comparison between task-resolving success rate and average costs per task for all tools.
 On average, \mytool takes 195 seconds and ~37k tokens (equivalent to 0.43 USD) to resolve one task instance in \swelite. In comparison, \sweagent costs 245k tokens (equivalent to 2.51 USD) per task instance. The cost of \devin is empty because it is not publicly available.
 When considering the combined three repetitions, \mytool takes 520 seconds (\textasciitilde 8.67 minutes) per task, which is below the 30-60 minute time limit deemed acceptable by developers for automated repair tools~\cite{icse22}.} 
 Looking into the 78 issues resolved by \mytool@3 in \swelite, it costs on average \textasciitilde 2.68 days for developers to create pull requests for 66 issues, and the other 12 issues take even longer to be closed by developers (ranging from 34 - 4023 days). The short response time and significantly low success-rate/cost ratio show the potential for \mytool to act as a first affordable step in future autonomous bug fixing.

\paragraph{Token Distribution of Resolved Tasks} \revised{We dived into the details of token cost distribution among resolved task instances by \mytool and \sweagent in Figure~\ref{fig:token-resolve}. The x-axis represents token cost ranges and the y-axis represents the number of resolved tasks. Figure~\ref{fig:token-resolve} shows that regarding the resolved tasks, \mytool is even more token-efficient. \mytool resolved 19 tasks with less than 10k tokens (\textasciitilde 0.1 USD) and 31 tasks between 10k-30k tokens, whereas 66.7\% of resolved tasks by \sweagent at least requires more than 100k tokens. This implies the strong capability of stratified context retrieval in \mytool to fast pinpoint error locations and thus produce final patch.}

\paragraph{Plausible / Correct patches.} Overfitting is a well-known challenge in the Automated Program Repair community~\cite{gao2019crash}. 
A program patch that passes the given test suite is said to be \emph{plausible}. 
However, a plausible patch is deemed as \emph{overfitting} if it fails to conform to the developer's intent. 
Otherwise, it is deemed as \emph{correct}.
To further understand the patch quality of \mytool and baselines, we manually verify the correctness of task-resolving (i.e. plausible) patches in \swelite.
Since three repetitions are performed, we consider a task to have a \emph{correct} patch if any of the three repetitions produced a \emph{correct} patch.
A plausible patch is correct if it is semantically equivalent to the developer patch.
In this verification process, at least two authors of the paper cross-validated each patch, and any disagreement was resolved with another author.
Overall, on \swelite, \mytool has a correctness rate of 65.4\% (51 correct/78 plausible).
We observed that the vast majority of \mytool's overfitting patches (all but 2 of the overfitting patches) modify the same methods as the developer patches, but the code modifications are wrong. This means that even the overfitting patches from \mytool are useful to the developer, since it helps in localization.
The main causes of wrong modifications are the limits of the LLM's capability or insufficient context. Apart from these, we noticed two other interesting causes of overfitting. One cause is that the issue creator gives a preliminary patch in the description. This patch can be different from the final developer patch, misleading the LLM. The other interesting cause is that the issue creator mentioned a case that needs to be handled. The LLM only fixes this mentioned case, but the developer fixed other similar cases as well. The two causes indicate that the issue description, just like test suites, can be an incomplete specification.

\subsection{RQ2: Effect of SBFL}
\label{sec:eval-rq2}

Bug reproduction~\cite{bugredux,bohme2017directed} is a well-studied topic in the software engineering community. It aims to automatically construct input tests to reproduce bugs that are described in bug reports. Subsequently, those bug reproduction tests can be used by program analysis techniques to localize the root cause of bugs and generate patches for APR tools~\cite{semfix}.

In this research question, we investigate whether program analysis techniques with reproducible tests can benefit the workflow of \mytool, and we use spectrum-based fault localization (SBFL) as an example.
Different from RQ1, here we construct a common scenario in program repair that \mytool has access to the complete test-suite of the target task instance.
We use the developer-written test cases for each task instance (provided in \swelite) as the test-suite.
To understand the effect of SBFL, we additionally provide SBFL results (top-5 suspicious methods) to \mytool at the beginning of the context retrieval stage, then use the test-suite for patch validation during the patch generation retry-loop. We denote this setting as \acr-sbfl in Table~\ref{tab:overall-result}. The patch validation is added as follows: when a patch is generated by the LLM agent, the test-suite is executed on the patched program; if the patch fails to pass the complete test-suite, \mytool re-invokes the patch generation agent to write a new patch. This validation loop is configured to run three times.



\paragraph{Results.} Table~\ref{tab:overall-result} shows that, with the additional information provided by SBFL, the number of resolved tasks increased from 57 to 66 (i.e. 19\% to 22\% resolved rate on \swelite).
Moreover, when comparing with task instances resolved in \mytool on \swelite, \acr-sbfl still uniquely resolves \textbf{7} task instances that are not resolved in any of the other runs.

\begin{figure}[t]
    \centering
    \includegraphics[width=0.95\columnwidth]{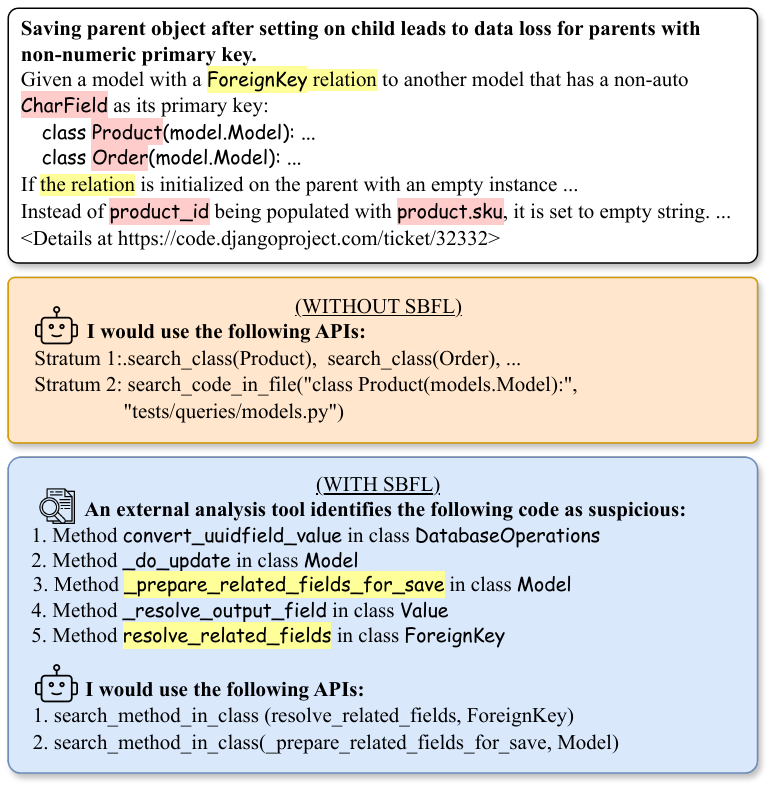}
    \caption{Issue description and \mytool's context retrieval (w. and w.o. SBFL), on django-13964.}
    \label{fig:sbfl-example}
\end{figure}

\paragraph{Case study.} We present a case study on one of the tasks uniquely resolved by \acr-sbfl.
Figure~\ref{fig:sbfl-example} shows the issue description of django-13964\footnote{\url{https://code.djangoproject.com/ticket/32332}} and how \mytool attempts to retrieve code context with and without SBFL.
This issue reported a bug when saving django models to a database.
A simplified version of the issue description is shown in the first part of Figure~\ref{fig:sbfl-example}, in which important parts are highlighted.
In this issue, some ``hints'' (highlighted in red) mentioned are actually distracting factors for a context retrieval agent.
For example, the \texttt{Product} and \texttt{Order} classes describe how the bug can be reproduced, and are not classes that cause the bug. Retrieving code context from these two classes (as shown in Figure~\ref{fig:sbfl-example} part 2 ``WITHOUT SBFL'') does not yield useful results for forming the context and resolving the issue.

On the other hand, the SBFL component can provide extra hints for context retrieval agent (as shown in Figure~\ref{fig:sbfl-example} part 3 ``WITH SBFL'').
This is because SBFL considers test execution differences, which in this case revealed a few more methods in the codebase that are related to the issue.
With these newly revealed hints, the agent decides to invoke APIs to search for the \texttt{resolve\_related\_fields} and \texttt{\_prepare\_related\_fields\_for\_save} methods (the latter\\ method is actually where the developer chose to fix this bug\footnote{\url{https://github.com/django/django/pull/13964/files}}).
Moreover, we observe that the agent does not solely rely on the SBFL results to make API invocations.
Instead of searching for methods ranked as top-1 in the SBFL results,
the agent searched for the 3rd and 5th ranked methods.
These methods are more related to some other hints mentioned in the issue description (highlighted in yellow), and the LLM agent is able to exploit this correlation between the natural language descriptions and the SBFL analysis results.
With the correct context collected, \mytool is then able to draft a patch that resolves this issue.
This suggests that an execution-based analysis can complement the agent workflow by revealing information not included in the issue description.





\subsection{RQ3: Challenges on real-life tasks}
\label{sec:eval-rq3}

In this research question, we analyze the task instances in \swelite that \mytool failed to resolve (based on the result of ACR-all in Table~\ref{tab:overall-result}, without applying SBFL) and provide a taxonomy of the issue characteristics to highlight the practical challenges in achieving fully automated software improvement. 
Our taxonomy consists of challenges in the fault localization stage and patch generation stage. 
Specifically, for each task, we analyze the best run in the three repetitions, and classify each of the 300 tasks into one of the following:

\begin{itemize}[leftmargin=1em]\itemsep0em 
    \item Success: The generated patch resolves the issue.
    \item Wrong patch: The generated patch modifies all methods that are modified in the developer patch. This means the patch content is wrong but the patch location(s) are correct.
    \item Wrong location in correct file: The generated patch that modifies the correct file but wrong location(s) in the file.
    \item Wrong file: The generated patch modifies the wrong file.
    \item No patch: No patch is generated from the retrieved context.
\end{itemize}

\begin{figure}[t]
    \centering    \includegraphics[width=0.85\columnwidth]{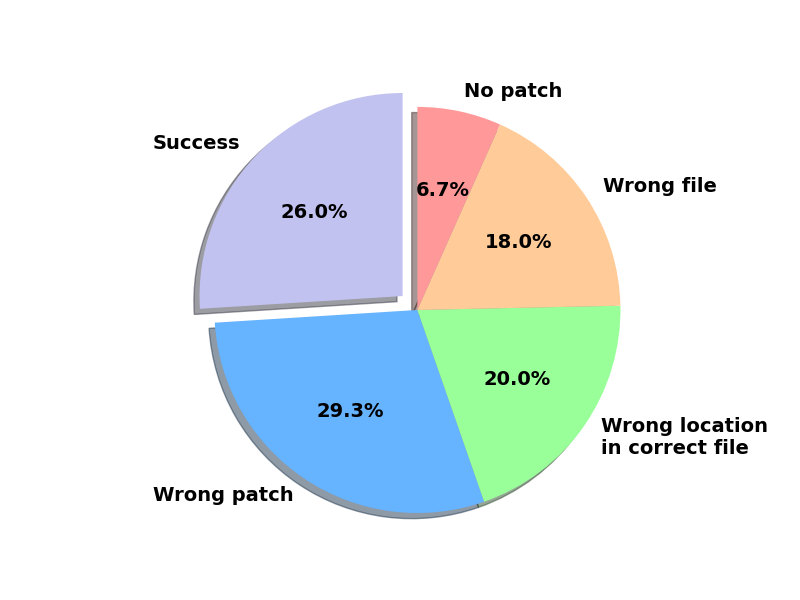}
    \vspace*{-0.2in}
    \caption{Taxonomy of Challenges in \swelite.}
    \label{fig:taxonomy}
\end{figure}

Figure~\ref{fig:taxonomy} shows the distribution of the 300 tasks in \swelite.
\mytool resolves 26.0\% of the issues (``Success''), as mentioned in Section~\ref{sec:eval-rq1}.
The fail-to-resolve cases are included in the remaining four categories.
In 29.3\% of the tasks, \mytool correctly decided on all patch locations (at the method-level), but did not produce a correct patch (``Wrong patch'').
More fine-grained intra-procedural analysis and specification inference techniques can play a significant role in improving these cases, by providing the patch generation agent with more method-level repair guidance. In the other three categories, the fault localization could not pinpoint all the locations to be modified.
In 20.0\% of the tasks, a patch is generated in the correct file, but at wrong methods / classes in the file (``Wrong location in correct file'').
In some of these runs, the developer patch modifies multiple methods, but the generated patch did not modify all of them.
In the other categories, a patch could not be generated at the correct file - in 18.0\% of the tasks a patch is generated in wrong files, and in 6.7\% of the tasks there is no applicable patch (``Wrong file'' and ``No patch'').
We manually inspected some tasks in these two categories, and observed that their issue description mentions few methods / classes / files in the codebase.
Instead, some of them contain short examples to reproduce the issue.
For these tasks, one possibility is to generate a comprehensive test-suite based on the issue description, and then use execution information of the test-suite (e.g. SBFL) to reveal suspicious program locations.
On the other hand, some other tasks do not contain reproducible examples and only consist of natural language descriptions. 
For these tasks some human involvement might be helpful. The developers could focus on these tasks. 






\section{Discussion on Experiments and Improvements}
\label{sec:discussion}



In this section, we discuss our position on the experiment results and a few possible directions for future improvements.

\begin{figure}[ht]
  \centering
    \includegraphics[width=\columnwidth]{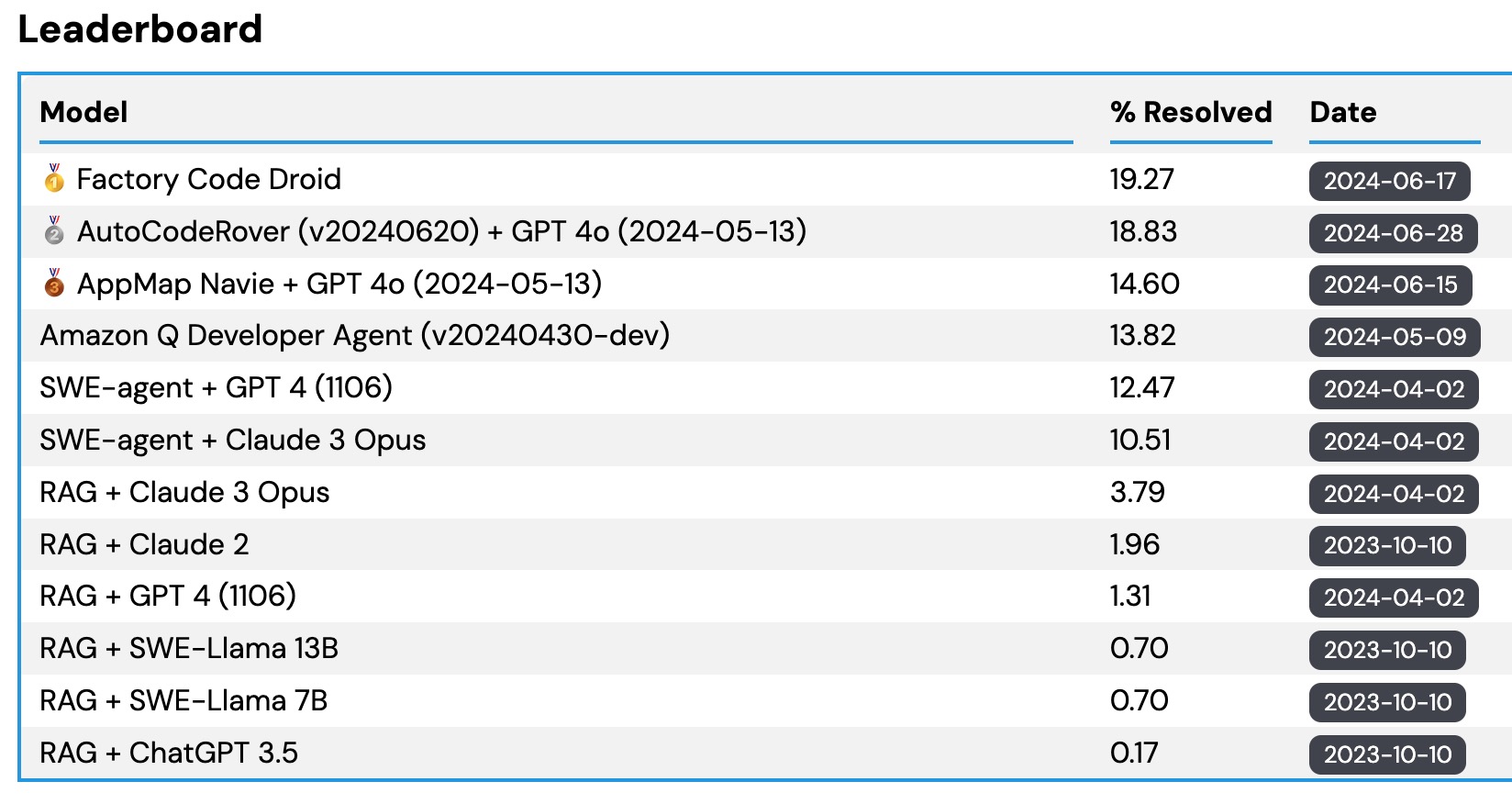}
  \caption{Snapshot of full \swe leaderboard on \snapshot.}
      \label{fig:swe-leaderboard}
\end{figure}

\paragraph{Position on Experiment Results} 
\revised{
Building autonomous large language model agent systems for software engineering tasks is one of the fastest-growing research fields now. There were more than 17 attempts from both academia and industries on \swe since April 2024. We refer the practitioners to the \swe Leaderboard~\cite{sweagentresult} (refer Figure~\ref{fig:swe-leaderboard}) which maintains the latest research effort of various recent agents. The most recent resolve rate on full \swe leaderboard has reached 19.27\% by Factory Code Droid as of \snapshot~. In the meantime, \mytool is also rapidly progressing towards better performance, the most recent update of \mytool (18.83\% on \swe on \snapshot~) can also be found at~\cite{autocoderover-web}. However, despite the satisfying
results of \mytool, it is useful to interpret the results in the right spirit. Apart from showing good efficacy with low cost, \mytool is significant because of the way it tries to generate program modifications. Thus, we highlight the importance of gleaning program
specifications to guide the patching process by agents as a general guideline for future research.}

\paragraph{Issue Reproducer} In Section~\ref{sec:methodology}, we described two scenarios that \mytool can work on, (1) issue description only, (2) aided by SBFL when the test suite is available. Although the second scenario is not always practical in real-life software development, the issue description sometimes contains a concrete reproduction script from the user that reports the bug. In the future, it is possible to design an LLM agent specifically to generate a bug reproduction test based on the GitHub issue description. The bug reproduction test can then be used to validate the correctness of \mytool's generated patch and possibly enable a regeneration process if a patch fails to pass the reproduction test. 

\paragraph{Semantic Artifacts} During context retrieval, \mytool effectively navigates through the codebase by visiting code entities such as classes and methods.
The idea of utilizing code-specific structure for context retrieval can be taken further by considering artifacts from program semantics.
For example, from an initial set of methods identified in the issue description, a static call graph analysis~\cite{originalcallgraph, pythoncallgraph} can be used to collect additional relevant methods when testcases are absent.
There are also potential in integrating a language server~\cite{languageserver} for codebase navigation, so that the context retrieval agent can perform more code-specific actions such as ``jump-to-definition'' from a method invocation. Finally, one can use forward data  dependence analysis \cite{wuxia24} from the methods in the issue description to search for other relevant methods.

\paragraph{Human Involvement} Currently, \mytool left all decision-making processes to its foundation LLM --- e.g. deciding context retrieving locations and whether to terminate the retrieval stage. However, leaving it to the LLM may not always be enough. Note that it is common to have multiple rounds of discussion even between the project maintainers before a pull request is created. Hence, a flexible interface and human involvement criteria between human developers and LLM agents are desired. 

\section{Threats to Validity}

Despite the high efficacy \mytool achieved in \swe, we report a few potential threats to our approach and experiments and discuss how we addressed them. 
First, LLM is known for its randomness to generate different results in different runs, which may threaten the validity of \mytool's performance. 
We address this by repeating our main evaluation experiment three times and releasing a replication package for practitioners.
Second, we consider a task instance in \swe as resolved if the generated patch passes all the test cases added in the pull request for the issue. 
However, this patch may be overfitting. 
We addressed this threat by manually verifying whether the patches are semantically equivalent to developers' patches among the authors.

\section{Perspectives}

AI-based software engineering is currently a topic of study among researchers, innovators, and entrepreneurs. Part of the trigger for this interest has been provided by efforts like GitHub Copilot \cite{gcopilot} in the last few years. These efforts show significant promise in the use of Large Language Models (LLMs) to automatically generate code. At the same time, code generated by LLMs may be incorrect \cite{fan2023automated} or vulnerable  \cite{karri23}. Thus, we need autonomous processes which allow for code improvements, such as bug fixes and feature additions.

In this paper, we suggest a LLM-based solution \mytool for autonomous code improvement. The key distinguishing feature of \mytool is its conscious attempt to inject a software engineering outlook by integrating (a) use of program representations such as ASTs instead of files, (b) iterative code search by exploiting program structure and (c) use of test-based fault localization when tests can be constructed. \mytool shows significant efficacy in terms of solving real-life GitHub issues. \mytool shows that relying only on the GitHub issue description to guide code search for patches / modifications can be misleading.  From the point of view of program repair, the toughest challenge lies in inferring the intent of the developer or the specification. The outlook of \mytool is that the developer intent is gleaned from the project structure for automated program modifications.

Today, the code LLMs cannot produce safe and secure code which can be trusted enough to be integrated into real software projects. There is thus a need to autonomously improve code (both automatically generated and manually written), for which LLMs can play a role.  Future work needs to focus further on the appropriate points when tools like \mytool may converse with the human programmer. Developers may need to shift to playing different roles at the same time in {\em future software industry} - vetting different conversations with LLM-based tools like \mytool to enable a variety of software engineering activities. This would contrast with {\em today's software industry} where a person has specific roles like programmer/tester/architect/requirements-engineer. Thus, apart from full-stack engineers, we may see more {\em full-lifecycle software engineers} in the future, who are comfortable working with the entire lifecycle of (the components of) a software system. Furthermore, the focus of LLM-oriented workflows is on {\em scale} today. This focus may shift to the engendering of {\em trust} if LLM-oriented autonomous software engineering becomes commonplace. Future software engineers may focus more on greater trust, instead of larger scale.

\section*{DATA AVAILABILITY}
We share full public access to (1) \mytool's implementation, (2) all patches generated during our experiment and conversation history with LLM, (3) replication scripts for our experiments at the following website: \link.
\begin{acks}

We thank the anonymous reviewers for their suggestions. This work was partially supported by a Singapore Ministry of Education (MoE) Tier 3 grant "Automated Program Repair", MOE-MOET32021-0001. 
\end{acks}


\onecolumn
\begin{multicols}{2}
    \bibliographystyle{ACM-Reference-Format}   
    \bibliography{references}


\begin{thebibliography}{48}


\ifx \showCODEN    \undefined \def \showCODEN     #1{\unskip}     \fi
\ifx \showDOI      \undefined \def \showDOI       #1{#1}\fi
\ifx \showISBNx    \undefined \def \showISBNx     #1{\unskip}     \fi
\ifx \showISBNxiii \undefined \def \showISBNxiii  #1{\unskip}     \fi
\ifx \showISSN     \undefined \def \showISSN      #1{\unskip}     \fi
\ifx \showLCCN     \undefined \def \showLCCN      #1{\unskip}     \fi
\ifx \shownote     \undefined \def \shownote      #1{#1}          \fi
\ifx \showarticletitle \undefined \def \showarticletitle #1{#1}   \fi
\ifx \showURL      \undefined \def \showURL       {\relax}        \fi
\providecommand\bibfield[2]{#2}
\providecommand\bibinfo[2]{#2}
\providecommand\natexlab[1]{#1}
\providecommand\showeprint[2][]{arXiv:#2}

\bibitem[gco(2022)]%
        {gcopilot}
 \bibinfo{year}{2022}\natexlab{}.
\newblock \bibinfo{title}{GitHub Copilot, your AI pair programmer}.
\newblock
\newblock
\urldef\tempurl%
\url{https://github.com/features/copilot/}
\showURL{%
\tempurl}


\bibitem[aut(2024)]%
        {autocoderover-web}
 \bibinfo{year}{2024}\natexlab{}.
\newblock \bibinfo{title}{AutoCodeRover, Autonomous Software Engineering}.
\newblock
\newblock
\urldef\tempurl%
\url{https://autocoderover.dev/}
\showURL{%
Retrieved July 10, 2024 from \tempurl}


\bibitem[Abreu et~al\mbox{.}(2007)]%
        {ochiai}
\bibfield{author}{\bibinfo{person}{Rui Abreu}, \bibinfo{person}{Peter Zoeteweij}, {and} \bibinfo{person}{Arjan~J.C. van Gemund}.} \bibinfo{year}{2007}\natexlab{}.
\newblock \showarticletitle{On the Accuracy of Spectrum-based Fault Localization}. In \bibinfo{booktitle}{\emph{Testing: Academic and Industrial Conference Practice and Research Techniques - MUTATION (TAICPART-MUTATION 2007)}}. IEEE, \bibinfo{publisher}{IEEE}, \bibinfo{pages}{89--98}.
\newblock
\urldef\tempurl%
\url{https://doi.org/10.1109/taic.part.2007.13}
\showDOI{\tempurl}


\bibitem[Bader et~al\mbox{.}(2019)]%
        {bader2019getafix}
\bibfield{author}{\bibinfo{person}{Johannes Bader}, \bibinfo{person}{Andrew Scott}, \bibinfo{person}{Michael Pradel}, {and} \bibinfo{person}{Satish Chandra}.} \bibinfo{year}{2019}\natexlab{}.
\newblock \showarticletitle{Getafix: Learning to fix bugs automatically}.
\newblock \bibinfo{journal}{\emph{Proceedings of the ACM on Programming Languages}} \bibinfo{volume}{3}, \bibinfo{number}{OOPSLA} (\bibinfo{year}{2019}), \bibinfo{pages}{1--27}.
\newblock


\bibitem[B{\"o}hme et~al\mbox{.}(2021)]%
        {bohme2020fuzzing}
\bibfield{author}{\bibinfo{person}{Marcel B{\"o}hme}, \bibinfo{person}{Cristian Cadar}, {and} \bibinfo{person}{Abhik Roychoudhury}.} \bibinfo{year}{2021}\natexlab{}.
\newblock \showarticletitle{Fuzzing: Challenges and Reflections}.
\newblock \bibinfo{journal}{\emph{IEEE Software}} \bibinfo{volume}{38}, \bibinfo{number}{3} (\bibinfo{year}{2021}).
\newblock


\bibitem[B{\"o}hme et~al\mbox{.}(2017)]%
        {bohme2017directed}
\bibfield{author}{\bibinfo{person}{Marcel B{\"o}hme}, \bibinfo{person}{Van-Thuan Pham}, \bibinfo{person}{Manh-Dung Nguyen}, {and} \bibinfo{person}{Abhik Roychoudhury}.} \bibinfo{year}{2017}\natexlab{}.
\newblock \showarticletitle{Directed greybox fuzzing}. In \bibinfo{booktitle}{\emph{Proceedings of the 2017 ACM SIGSAC conference on computer and communications security}}. \bibinfo{pages}{2329--2344}.
\newblock


\bibitem[Cadar and Sen(2013)]%
        {cadar-sen}
\bibfield{author}{\bibinfo{person}{Cristian Cadar} {and} \bibinfo{person}{Koushik Sen}.} \bibinfo{year}{2013}\natexlab{}.
\newblock \showarticletitle{Symbolic execution for software testing: three decades later}.
\newblock \bibinfo{journal}{\emph{Commun. ACM}} \bibinfo{volume}{56}, \bibinfo{number}{2} (\bibinfo{year}{2013}).
\newblock


\bibitem[Chen et~al\mbox{.}(2021)]%
        {codex}
\bibfield{author}{\bibinfo{person}{Mark Chen}, \bibinfo{person}{Jerry Tworek}, \bibinfo{person}{Heewoo Jun}, \bibinfo{person}{Qiming Yuan}, \bibinfo{person}{Henrique Pondé~de Oliveira~Pinto}, \bibinfo{person}{Jared Kaplan}, \bibinfo{person}{Harrison Edwards}, \bibinfo{person}{Yuri Burda}, \bibinfo{person}{Nicholas Joseph}, \bibinfo{person}{Greg Brockman}, \bibinfo{person}{Alex Ray}, \bibinfo{person}{Raul Puri}, \bibinfo{person}{Gretchen Krueger}, \bibinfo{person}{Michael Petrov}, \bibinfo{person}{Heidy Khlaaf}, \bibinfo{person}{Girish Sastry}, \bibinfo{person}{Pamela Mishkin}, \bibinfo{person}{Brooke Chan}, \bibinfo{person}{Scott Gray}, \bibinfo{person}{Nick Ryder}, \bibinfo{person}{Mikhail Pavlov}, \bibinfo{person}{Alethea Power}, \bibinfo{person}{Lukasz Kaiser}, \bibinfo{person}{Mohammad Bavarian}, \bibinfo{person}{Clemens Winter}, \bibinfo{person}{Philippe Tillet}, \bibinfo{person}{Felipe~Petroski Such}, \bibinfo{person}{Dave Cummings}, \bibinfo{person}{Matthias Plappert}, \bibinfo{person}{Fotios
  Chantzis}, \bibinfo{person}{Elizabeth Barnes}, \bibinfo{person}{Ariel Herbert-Voss}, \bibinfo{person}{William~Hebgen Guss}, \bibinfo{person}{Alex Nichol}, \bibinfo{person}{Alex Paino}, \bibinfo{person}{Nikolas Tezak}, \bibinfo{person}{Jie Tang}, \bibinfo{person}{Igor Babuschkin}, \bibinfo{person}{Suchir Balaji}, \bibinfo{person}{Shantanu Jain}, \bibinfo{person}{William Saunders}, \bibinfo{person}{Christopher Hesse}, \bibinfo{person}{Andrew~N. Carr}, \bibinfo{person}{Jan Leike}, \bibinfo{person}{Joshua Achiam}, \bibinfo{person}{Vedant Misra}, \bibinfo{person}{Evan Morikawa}, \bibinfo{person}{Alec Radford}, \bibinfo{person}{Matthew Knight}, \bibinfo{person}{Miles Brundage}, \bibinfo{person}{Mira Murati}, \bibinfo{person}{Katie Mayer}, \bibinfo{person}{Peter Welinder}, \bibinfo{person}{Bob McGrew}, \bibinfo{person}{Dario Amodei}, \bibinfo{person}{Sam McCandlish}, \bibinfo{person}{Ilya Sutskever}, {and} \bibinfo{person}{Wojciech Zaremba}.} \bibinfo{year}{2021}\natexlab{}.
\newblock \showarticletitle{Evaluating Large Language Models Trained on Code.}
\newblock \bibinfo{journal}{\emph{arXiv.org}}  \bibinfo{volume}{abs/2107.03374} (\bibinfo{date}{7} \bibinfo{year}{2021}).
\newblock
\showISSN{2331-8422}
\showeprint[arXiv]{2107.03374}
\urldef\tempurl%
\url{https://arxiv.org/abs/2107.03374}
\showURL{%
\tempurl}


\bibitem[Chow et~al\mbox{.}(2024)]%
        {chow2024pyty}
\bibfield{author}{\bibinfo{person}{Yiu~Wai Chow}, \bibinfo{person}{Luca Di~Grazia}, {and} \bibinfo{person}{Michael Pradel}.} \bibinfo{year}{2024}\natexlab{}.
\newblock \showarticletitle{Pyty: Repairing static type errors in python}. In \bibinfo{booktitle}{\emph{Proceedings of the IEEE/ACM 46th International Conference on Software Engineering}}. \bibinfo{pages}{1--13}.
\newblock


\bibitem[Fan et~al\mbox{.}(2023)]%
        {fan2023automated}
\bibfield{author}{\bibinfo{person}{Zhiyu Fan}, \bibinfo{person}{Xiang Gao}, \bibinfo{person}{Martin Mirchev}, \bibinfo{person}{Abhik Roychoudhury}, {and} \bibinfo{person}{Shin~Hwei Tan}.} \bibinfo{year}{2023}\natexlab{}.
\newblock \showarticletitle{Automated Repair of Programs from Large Language Models.}. In \bibinfo{booktitle}{\emph{45th IEEE/ACM International Conference on Software Engineering, ICSE 2023, Melbourne, Australia, May 14-20, 2023}}. IEEE, \bibinfo{publisher}{IEEE}, \bibinfo{pages}{1469--1481}.
\newblock
\urldef\tempurl%
\url{https://doi.org/10.1109/icse48619.2023.00128}
\showDOI{\tempurl}


\bibitem[Fu et~al\mbox{.}(2022)]%
        {fu2022vulrepair}
\bibfield{author}{\bibinfo{person}{Michael Fu}, \bibinfo{person}{Chakkrit Tantithamthavorn}, \bibinfo{person}{Trung Le}, \bibinfo{person}{Van Nguyen}, {and} \bibinfo{person}{Dinh Phung}.} \bibinfo{year}{2022}\natexlab{}.
\newblock \showarticletitle{VulRepair: a T5-based automated software vulnerability repair}. In \bibinfo{booktitle}{\emph{Proceedings of the 30th ACM joint european software engineering conference and symposium on the foundations of software engineering}}. \bibinfo{pages}{935--947}.
\newblock


\bibitem[Gao et~al\mbox{.}(2019)]%
        {gao2019crash}
\bibfield{author}{\bibinfo{person}{Xiang Gao}, \bibinfo{person}{Sergey Mechtaev}, {and} \bibinfo{person}{Abhik Roychoudhury}.} \bibinfo{year}{2019}\natexlab{}.
\newblock \showarticletitle{Crash-avoiding program repair}. In \bibinfo{booktitle}{\emph{Proceedings of the 28th ACM SIGSOFT International Symposium on Software Testing and Analysis}}. \bibinfo{pages}{8--18}.
\newblock


\bibitem[Goues et~al\mbox{.}(2019)]%
        {cacm19}
\bibfield{author}{\bibinfo{person}{Claire~Le Goues}, \bibinfo{person}{Michael Pradel}, {and} \bibinfo{person}{Abhik Roychoudhury}.} \bibinfo{year}{2019}\natexlab{}.
\newblock \showarticletitle{Automated program repair.}
\newblock \bibinfo{journal}{\emph{Commun. ACM}}  \bibinfo{volume}{62} (\bibinfo{date}{11} \bibinfo{year}{2019}), \bibinfo{pages}{56--65}.
\newblock
Issue 12.
\showISSN{0001-0782}


\bibitem[Gunasinghe and Marcus(2021)]%
        {languageserver}
\bibfield{author}{\bibinfo{person}{Nadeeshaan Gunasinghe} {and} \bibinfo{person}{Nipuna Marcus}.} \bibinfo{year}{2021}\natexlab{}.
\newblock \bibinfo{booktitle}{\emph{Language Server Protocol and Implementation}}.
\newblock \bibinfo{publisher}{Springer}.
\newblock


\bibitem[Jiang et~al\mbox{.}(2023)]%
        {jiang2023impact}
\bibfield{author}{\bibinfo{person}{Nan Jiang}, \bibinfo{person}{Kevin Liu}, \bibinfo{person}{Thibaud Lutellier}, {and} \bibinfo{person}{Lin Tan}.} \bibinfo{year}{2023}\natexlab{}.
\newblock \showarticletitle{Impact of Code Language Models on Automated Program Repair.}, In \bibinfo{booktitle}{45th IEEE/ACM International Conference on Software Engineering, ICSE 2023, Melbourne, Australia, May 14-20, 2023} (Melbourne, Victoria, Australia).
\newblock \bibinfo{journal}{\emph{International Conference on Software Engineering}}, \bibinfo{pages}{1430--1442}.
\newblock
\showISBNx{9781665457019}
\urldef\tempurl%
\url{https://doi.org/10.1109/icse48619.2023.00125}
\showDOI{\tempurl}


\bibitem[Jimenez et~al\mbox{.}(2024a)]%
        {sweagentresult}
\bibfield{author}{\bibinfo{person}{Carlos~E. Jimenez}, \bibinfo{person}{John Yang}, \bibinfo{person}{Alexander Wettig}, \bibinfo{person}{Shunyu Yao}, \bibinfo{person}{Kexin Pei}, \bibinfo{person}{Ofir Press}, {and} \bibinfo{person}{Karthik Narasimhan}.} \bibinfo{year}{2024}\natexlab{a}.
\newblock \bibinfo{title}{Leaderboard results on SWE-bench}.
\newblock
\newblock
\urldef\tempurl%
\url{https://www.swebench.com/}
\showURL{%
Retrieved April 8, 2024 from \tempurl}


\bibitem[Jimenez et~al\mbox{.}(2024b)]%
        {jimenez2024swebench}
\bibfield{author}{\bibinfo{person}{Carlos~E Jimenez}, \bibinfo{person}{John Yang}, \bibinfo{person}{Alexander Wettig}, \bibinfo{person}{Shunyu Yao}, \bibinfo{person}{Kexin Pei}, \bibinfo{person}{Ofir Press}, {and} \bibinfo{person}{Karthik~R Narasimhan}.} \bibinfo{year}{2024}\natexlab{b}.
\newblock \showarticletitle{{SWE}-bench: Can Language Models Resolve Real-world Github Issues?}. In \bibinfo{booktitle}{\emph{The Twelfth International Conference on Learning Representations}}.
\newblock
\urldef\tempurl%
\url{https://openreview.net/forum?id=VTF8yNQM66}
\showURL{%
\tempurl}


\bibitem[Jin et~al\mbox{.}(2023)]%
        {jin2023inferfix}
\bibfield{author}{\bibinfo{person}{Matthew Jin}, \bibinfo{person}{Syed Shahriar}, \bibinfo{person}{Michele Tufano}, \bibinfo{person}{Xin Shi}, \bibinfo{person}{Shuai Lu}, \bibinfo{person}{Neel Sundaresan}, {and} \bibinfo{person}{Alexey Svyatkovskiy}.} \bibinfo{year}{2023}\natexlab{}.
\newblock \showarticletitle{Inferfix: End-to-end program repair with llms}. In \bibinfo{booktitle}{\emph{Proceedings of the 31st ACM Joint European Software Engineering Conference and Symposium on the Foundations of Software Engineering}}. \bibinfo{pages}{1646--1656}.
\newblock


\bibitem[Jin et~al\mbox{.}(2024)]%
        {wuxia24}
\bibfield{author}{\bibinfo{person}{Wuxia Jin} {et~al\mbox{.}}} \bibinfo{year}{2024}\natexlab{}.
\newblock \showarticletitle{PyAnalyzer: An Effective and Practical Approach for Dependency Extraction from Python Code}. In \bibinfo{booktitle}{\emph{International Conference on Software Engineering (ICSE)}}.
\newblock


\bibitem[Jin and Orso(2012)]%
        {bugredux}
\bibfield{author}{\bibinfo{person}{Wei Jin} {and} \bibinfo{person}{Alessandro Orso}.} \bibinfo{year}{2012}\natexlab{}.
\newblock \showarticletitle{BugRedux: Reproducing field failures for in-house debugging}. In \bibinfo{booktitle}{\emph{2012 34th International Conference on Software Engineering (ICSE)}}. \bibinfo{pages}{474--484}.
\newblock
\urldef\tempurl%
\url{https://doi.org/10.1109/ICSE.2012.6227168}
\showDOI{\tempurl}


\bibitem[Jones et~al\mbox{.}(2002)]%
        {tarantula}
\bibfield{author}{\bibinfo{person}{James~A Jones}, \bibinfo{person}{Mary~Jean Harrold}, {and} \bibinfo{person}{John Stasko}.} \bibinfo{year}{2002}\natexlab{}.
\newblock \showarticletitle{Visualization of test information to assist fault localization}. In \bibinfo{booktitle}{\emph{Proceedings of the 24th international conference on Software engineering}}. \bibinfo{pages}{467--477}.
\newblock


\bibitem[Keller et~al\mbox{.}(2017)]%
        {sbfleval}
\bibfield{author}{\bibinfo{person}{Fabian Keller}, \bibinfo{person}{Lars Grunske}, \bibinfo{person}{Simon Heiden}, \bibinfo{person}{Antonio Filieri}, \bibinfo{person}{Andre van Hoorn}, {and} \bibinfo{person}{David Lo}.} \bibinfo{year}{2017}\natexlab{}.
\newblock \showarticletitle{A critical evaluation of spectrum-based fault localization techniques on a large-scale software system}. In \bibinfo{booktitle}{\emph{2017 IEEE International Conference on Software Quality, Reliability and Security (QRS)}}. IEEE, \bibinfo{pages}{114--125}.
\newblock


\bibitem[Koyuncu et~al\mbox{.}(2019)]%
        {koyuncu2019ifixr}
\bibfield{author}{\bibinfo{person}{Anil Koyuncu}, \bibinfo{person}{Kui Liu}, \bibinfo{person}{Tegawend{\'e}~F Bissyand{\'e}}, \bibinfo{person}{Dongsun Kim}, \bibinfo{person}{Martin Monperrus}, \bibinfo{person}{Jacques Klein}, {and} \bibinfo{person}{Yves Le~Traon}.} \bibinfo{year}{2019}\natexlab{}.
\newblock \showarticletitle{iFixR: Bug report driven program repair}. In \bibinfo{booktitle}{\emph{Proceedings of the 2019 27th ACM joint meeting on european software engineering conference and symposium on the foundations of software engineering}}. \bibinfo{pages}{314--325}.
\newblock


\bibitem[Labs(2024)]%
        {devin}
\bibfield{author}{\bibinfo{person}{Cognition Labs}.} \bibinfo{year}{2024}\natexlab{}.
\newblock \bibinfo{title}{Devin, AI software engineer}.
\newblock
\newblock
\urldef\tempurl%
\url{https://www.cognition-labs.com/introducing-devin}
\showURL{%
Retrieved April 12, 2024 from \tempurl}


\bibitem[Li et~al\mbox{.}(2022)]%
        {alphacode}
\bibfield{author}{\bibinfo{person}{Yujia Li}, \bibinfo{person}{David~H. Choi}, \bibinfo{person}{Junyoung Chung}, \bibinfo{person}{Nate Kushman}, \bibinfo{person}{Julian Schrittwieser}, \bibinfo{person}{Rémi Leblond}, \bibinfo{person}{Tom Eccles}, \bibinfo{person}{James Keeling}, \bibinfo{person}{Felix Gimeno}, \bibinfo{person}{Agustin~Dal Lago}, \bibinfo{person}{Thomas Hubert}, \bibinfo{person}{Peter Choy}, \bibinfo{person}{Cyprien~de Masson~d'Autume}, \bibinfo{person}{Igor Babuschkin}, \bibinfo{person}{Xinyun Chen}, \bibinfo{person}{Po-Sen Huang}, \bibinfo{person}{Johannes Welbl}, \bibinfo{person}{Sven Gowal}, \bibinfo{person}{Alexey Cherepanov}, \bibinfo{person}{James Molloy}, \bibinfo{person}{Daniel~J. Mankowitz}, \bibinfo{person}{Esme~Sutherland Robson}, \bibinfo{person}{Pushmeet Kohli}, \bibinfo{person}{Nando de Freitas}, \bibinfo{person}{Koray Kavukcuoglu}, {and} \bibinfo{person}{Oriol Vinyals}.} \bibinfo{year}{2022}\natexlab{}.
\newblock \showarticletitle{Competition-Level Code Generation with AlphaCode.}
\newblock \bibinfo{journal}{\emph{Science}} \bibinfo{volume}{abs/2203.07814}, \bibinfo{number}{6624} (\bibinfo{date}{12} \bibinfo{year}{2022}), \bibinfo{pages}{1092--1097}.
\newblock
\showISSN{0036-8075}
\urldef\tempurl%
\url{https://doi.org/10.48550/arxiv.2203.07814}
\showDOI{\tempurl}


\bibitem[Lutellier et~al\mbox{.}(2020)]%
        {lutellier2020coconut}
\bibfield{author}{\bibinfo{person}{Thibaud Lutellier}, \bibinfo{person}{Hung~Viet Pham}, \bibinfo{person}{Lawrence Pang}, \bibinfo{person}{Yitong Li}, \bibinfo{person}{Moshi Wei}, {and} \bibinfo{person}{Lin Tan}.} \bibinfo{year}{2020}\natexlab{}.
\newblock \showarticletitle{CoCoNuT: combining context-aware neural translation models using ensemble for program repair.}, In \bibinfo{booktitle}{ISSTA '20: 29th ACM SIGSOFT International Symposium on Software Testing and Analysis, Virtual Event, USA, July 18-22, 2020}, \bibfield{editor}{\bibinfo{person}{Sarfraz Khurshid} {and} \bibinfo{person}{Corina~S. Pasareanu}} (Eds.).
\newblock \bibinfo{journal}{\emph{International Symposium on Software Testing and Analysis}}, \bibinfo{pages}{101--114}.
\newblock


\bibitem[Marginean et~al\mbox{.}(2019)]%
        {marginean2019sapfix}
\bibfield{author}{\bibinfo{person}{Alexandru Marginean}, \bibinfo{person}{Johannes Bader}, \bibinfo{person}{Satish Chandra}, \bibinfo{person}{Mark Harman}, \bibinfo{person}{Yue Jia}, \bibinfo{person}{Ke Mao}, \bibinfo{person}{Alexander Mols}, {and} \bibinfo{person}{Andrew Scott}.} \bibinfo{year}{2019}\natexlab{}.
\newblock \showarticletitle{Sapfix: Automated end-to-end repair at scale}. In \bibinfo{booktitle}{\emph{2019 IEEE/ACM 41st International Conference on Software Engineering: Software Engineering in Practice (ICSE-SEIP)}}. IEEE, \bibinfo{pages}{269--278}.
\newblock


\bibitem[Mechtaev et~al\mbox{.}(2016)]%
        {angelix}
\bibfield{author}{\bibinfo{person}{Sergey Mechtaev}, \bibinfo{person}{Jooyong Yi}, {and} \bibinfo{person}{Abhik Roychoudhury}.} \bibinfo{year}{2016}\natexlab{}.
\newblock \showarticletitle{Angelix: scalable multiline program patch synthesis via symbolic analysis.}, In \bibinfo{booktitle}{Proceedings of the 38th International Conference on Software Engineering, ICSE 2016, Austin, TX, USA, May 14-22, 2016}, \bibfield{editor}{\bibinfo{person}{Laura~K. Dillon}, \bibinfo{person}{Willem Visser}, {and} \bibinfo{person}{Laurie Williams}} (Eds.).
\newblock \bibinfo{journal}{\emph{International Conference on Software Engineering}}, \bibinfo{pages}{691--701}.
\newblock


\bibitem[Motwani and Brun(2023)]%
        {motwani2023better}
\bibfield{author}{\bibinfo{person}{Manish Motwani} {and} \bibinfo{person}{Yuriy Brun}.} \bibinfo{year}{2023}\natexlab{}.
\newblock \showarticletitle{Better automatic program repair by using bug reports and tests together}. In \bibinfo{booktitle}{\emph{2023 IEEE/ACM 45th International Conference on Software Engineering (ICSE)}}. IEEE, \bibinfo{pages}{1225--1237}.
\newblock


\bibitem[Nguyen et~al\mbox{.}(2013)]%
        {semfix}
\bibfield{author}{\bibinfo{person}{Hoang Duong~Thien Nguyen}, \bibinfo{person}{Dawei Qi}, \bibinfo{person}{Abhik Roychoudhury}, {and} \bibinfo{person}{Satish Chandra}.} \bibinfo{year}{2013}\natexlab{}.
\newblock \showarticletitle{SemFix: program repair via semantic analysis.}, In \bibinfo{booktitle}{35th International Conference on Software Engineering, ICSE '13, San Francisco, CA, USA, May 18-26, 2013} (San Francisco, CA, USA), \bibfield{editor}{\bibinfo{person}{David Notkin}, \bibinfo{person}{Betty H.~C. Cheng}, {and} \bibinfo{person}{Klaus Pohl}} (Eds.).
\newblock \bibinfo{journal}{\emph{International Conference on Software Engineering}}, \bibinfo{pages}{772--781}.
\newblock
\showISBNx{978-1-4673-3076-3}
\urldef\tempurl%
\url{https://doi.org/10.1109/icse.2013.6606623}
\showDOI{\tempurl}


\bibitem[Noller et~al\mbox{.}(2022)]%
        {icse22}
\bibfield{author}{\bibinfo{person}{Yannic Noller}, \bibinfo{person}{Ridwan Shariffdeen}, \bibinfo{person}{Xiang Gao}, {and} \bibinfo{person}{Abhik Roychoudhury}.} \bibinfo{year}{2022}\natexlab{}.
\newblock \showarticletitle{Trust Enhancement Issues in Program Repair}. In \bibinfo{booktitle}{\emph{IEEE/ACM 44th International Conference on Software Engineering (ICSE)}}.
\newblock


\bibitem[Ovalle(2023)]%
        {copilot}
\bibfield{author}{\bibinfo{person}{Brayan Stiven~Torrres Ovalle}.} \bibinfo{year}{2023}\natexlab{}.
\newblock \bibinfo{title}{GitHub Copilot}.
\newblock
\newblock
\urldef\tempurl%
\url{https://doi.org/10.26507/paper.2300}
\showDOI{\tempurl}


\bibitem[Pearce et~al\mbox{.}(2023a)]%
        {pearce2023examining}
\bibfield{author}{\bibinfo{person}{Hammond Pearce}, \bibinfo{person}{Benjamin Tan}, \bibinfo{person}{Baleegh Ahmad}, \bibinfo{person}{Ramesh Karri}, {and} \bibinfo{person}{Brendan Dolan-Gavitt}.} \bibinfo{year}{2023}\natexlab{a}.
\newblock \showarticletitle{Examining zero-shot vulnerability repair with large language models}. In \bibinfo{booktitle}{\emph{2023 IEEE Symposium on Security and Privacy (SP)}}. IEEE, \bibinfo{pages}{2339--2356}.
\newblock


\bibitem[Pearce et~al\mbox{.}(2023b)]%
        {karri23}
\bibfield{author}{\bibinfo{person}{H Pearce}, \bibinfo{person}{B Tan}, \bibinfo{person}{B Ahmad}, \bibinfo{person}{R Karri}, {and} \bibinfo{person}{B Dolan-Gavitt}.} \bibinfo{year}{2023}\natexlab{b}.
\newblock \showarticletitle{Examining zero-shot vulnerability repair with large language models}. In \bibinfo{booktitle}{\emph{IEEE Symposium on Security and Privacy (SP)}}.
\newblock


\bibitem[Prenner et~al\mbox{.}(2022)]%
        {prenner2022can}
\bibfield{author}{\bibinfo{person}{Julian~Aron Prenner}, \bibinfo{person}{Hlib Babii}, {and} \bibinfo{person}{Romain Robbes}.} \bibinfo{year}{2022}\natexlab{}.
\newblock \showarticletitle{Can OpenAI's codex fix bugs? an evaluation on QuixBugs}. In \bibinfo{booktitle}{\emph{Proceedings of the Third International Workshop on Automated Program Repair}}. \bibinfo{pages}{69--75}.
\newblock


\bibitem[Ryder(1979)]%
        {originalcallgraph}
\bibfield{author}{\bibinfo{person}{Barbara~G Ryder}.} \bibinfo{year}{1979}\natexlab{}.
\newblock \showarticletitle{Constructing the call graph of a program}.
\newblock \bibinfo{journal}{\emph{IEEE Transactions on Software Engineering}} \bibinfo{number}{3} (\bibinfo{year}{1979}), \bibinfo{pages}{216--226}.
\newblock


\bibitem[Salis et~al\mbox{.}(2021)]%
        {pythoncallgraph}
\bibfield{author}{\bibinfo{person}{Vitalis Salis}, \bibinfo{person}{Thodoris Sotiropoulos}, \bibinfo{person}{Panos Louridas}, \bibinfo{person}{Diomidis Spinellis}, {and} \bibinfo{person}{Dimitris Mitropoulos}.} \bibinfo{year}{2021}\natexlab{}.
\newblock \showarticletitle{Pycg: Practical call graph generation in python}. In \bibinfo{booktitle}{\emph{2021 IEEE/ACM 43rd International Conference on Software Engineering (ICSE)}}. IEEE, \bibinfo{pages}{1646--1657}.
\newblock


\bibitem[Smith et~al\mbox{.}(2015)]%
        {cure}
\bibfield{author}{\bibinfo{person}{Edward~K. Smith}, \bibinfo{person}{Earl~T. Barr}, \bibinfo{person}{Claire~Le Goues}, {and} \bibinfo{person}{Yuriy Brun}.} \bibinfo{year}{2015}\natexlab{}.
\newblock \showarticletitle{Is the cure worse than the disease? overfitting in automated program repair.}, In \bibinfo{booktitle}{Proceedings of the 2015 10th Joint Meeting on Foundations of Software Engineering, ESEC/FSE 2015, Bergamo, Italy, August 30 - September 4, 2015}, \bibfield{editor}{\bibinfo{person}{Elisabetta~Di Nitto}, \bibinfo{person}{Mark Harman}, {and} \bibinfo{person}{Patrick Heymans}} (Eds.).
\newblock \bibinfo{journal}{\emph{ESEC/SIGSOFT FSE}}, \bibinfo{pages}{532--543}.
\newblock
\urldef\tempurl%
\url{http://people.cs.umass.edu/%7Ebrun/pubs/pubs/Smith15fse.pdf}
\showURL{%
\tempurl}


\bibitem[Tan et~al\mbox{.}(2024)]%
        {tan2024crossfix}
\bibfield{author}{\bibinfo{person}{Shin~Hwei Tan}, \bibinfo{person}{Ziqiang Li}, {and} \bibinfo{person}{Lu Yan}.} \bibinfo{year}{2024}\natexlab{}.
\newblock \showarticletitle{CrossFix: Resolution of GitHub issues via similar bugs recommendation}.
\newblock \bibinfo{journal}{\emph{Journal of Software: Evolution and Process}} \bibinfo{volume}{36}, \bibinfo{number}{4} (\bibinfo{year}{2024}), \bibinfo{pages}{e2554}.
\newblock


\bibitem[Team(2024)]%
        {devinreport}
\bibfield{author}{\bibinfo{person}{The~Cognition Team}.} \bibinfo{year}{2024}\natexlab{}.
\newblock \bibinfo{title}{SWE-bench technical report (Devin)}.
\newblock
\newblock
\urldef\tempurl%
\url{https://www.cognition-labs.com/post/swe-bench-technical-report}
\showURL{%
Retrieved April 12, 2024 from \tempurl}


\bibitem[Wang et~al\mbox{.}(2023)]%
        {wang2023rap}
\bibfield{author}{\bibinfo{person}{Weishi Wang}, \bibinfo{person}{Yue Wang}, \bibinfo{person}{Shafiq Joty}, {and} \bibinfo{person}{Steven~CH Hoi}.} \bibinfo{year}{2023}\natexlab{}.
\newblock \showarticletitle{Rap-gen: Retrieval-augmented patch generation with codet5 for automatic program repair}. In \bibinfo{booktitle}{\emph{Proceedings of the 31st ACM Joint European Software Engineering Conference and Symposium on the Foundations of Software Engineering}}. \bibinfo{pages}{146--158}.
\newblock


\bibitem[Weimer et~al\mbox{.}(2009)]%
        {genprog}
\bibfield{author}{\bibinfo{person}{Westley Weimer}, \bibinfo{person}{ThanhVu Nguyen}, \bibinfo{person}{Claire~Le Goues}, {and} \bibinfo{person}{Stephanie Forrest}.} \bibinfo{year}{2009}\natexlab{}.
\newblock \showarticletitle{Automatically finding patches using genetic programming.}, In \bibinfo{booktitle}{31st International Conference on Software Engineering, ICSE 2009, May 16-24, 2009, Vancouver, Canada, Proceedings}.
\newblock \bibinfo{journal}{\emph{2009 IEEE 31st International Conference on Software Engineering}}, \bibinfo{pages}{364--374}.
\newblock
\urldef\tempurl%
\url{https://doi.org/10.1109/icse.2009.5070536}
\showDOI{\tempurl}


\bibitem[Williams et~al\mbox{.}(2023)]%
        {williams2023user}
\bibfield{author}{\bibinfo{person}{David Williams}, \bibinfo{person}{James Callan}, \bibinfo{person}{Serkan Kirbas}, \bibinfo{person}{Sergey Mechtaev}, \bibinfo{person}{Justyna Petke}, \bibinfo{person}{Thomas Prideaux-Ghee}, {and} \bibinfo{person}{Federica Sarro}.} \bibinfo{year}{2023}\natexlab{}.
\newblock \showarticletitle{User-Centric Deployment of Automated Program Repair at Bloomberg}.
\newblock \bibinfo{journal}{\emph{arXiv preprint arXiv:2311.10516}} (\bibinfo{year}{2023}).
\newblock


\bibitem[Wong et~al\mbox{.}(2016a)]%
        {sbfl-survey}
\bibfield{author}{\bibinfo{person}{WE Wong}, \bibinfo{person}{R Gao}, \bibinfo{person}{Y Li}, \bibinfo{person}{R Abreu}, {and} \bibinfo{person}{F Wotawa}.} \bibinfo{year}{2016}\natexlab{a}.
\newblock \showarticletitle{A survey on software fault localization}.
\newblock \bibinfo{journal}{\emph{IEEE Transactions on Software Engineering}} (\bibinfo{year}{2016}), \bibinfo{pages}{707--740}.
\newblock
Issue 8.


\bibitem[Wong et~al\mbox{.}(2016b)]%
        {flsurvey}
\bibfield{author}{\bibinfo{person}{W~Eric Wong}, \bibinfo{person}{Ruizhi Gao}, \bibinfo{person}{Yihao Li}, \bibinfo{person}{Rui Abreu}, {and} \bibinfo{person}{Franz Wotawa}.} \bibinfo{year}{2016}\natexlab{b}.
\newblock \showarticletitle{A survey on software fault localization}.
\newblock \bibinfo{journal}{\emph{IEEE Transactions on Software Engineering}} \bibinfo{volume}{42}, \bibinfo{number}{8} (\bibinfo{year}{2016}), \bibinfo{pages}{707--740}.
\newblock


\bibitem[Xia et~al\mbox{.}(2023)]%
        {xia2023automated}
\bibfield{author}{\bibinfo{person}{Chunqiu~Steven Xia}, \bibinfo{person}{Yuxiang Wei}, {and} \bibinfo{person}{Lingming Zhang}.} \bibinfo{year}{2023}\natexlab{}.
\newblock \showarticletitle{Automated program repair in the era of large pre-trained language models}. In \bibinfo{booktitle}{\emph{2023 IEEE/ACM 45th International Conference on Software Engineering (ICSE)}}. IEEE, \bibinfo{pages}{1482--1494}.
\newblock


\bibitem[Yang et~al\mbox{.}(2024)]%
        {yang2024sweagent}
\bibfield{author}{\bibinfo{person}{John Yang}, \bibinfo{person}{Carlos~E. Jimenez}, \bibinfo{person}{Alexander Wettig}, \bibinfo{person}{Kilian Lieret}, \bibinfo{person}{Shunyu Yao}, \bibinfo{person}{Karthik Narasimhan}, {and} \bibinfo{person}{Ofir Press}.} \bibinfo{year}{2024}\natexlab{}.
\newblock \bibinfo{title}{SWE-agent: Agent-Computer Interfaces Enable Automated Software Engineering}.
\newblock
\newblock
\showeprint[arxiv]{2405.15793}~[cs.SE]


\bibitem[Zhu et~al\mbox{.}(2021)]%
        {recoder}
\bibfield{author}{\bibinfo{person}{Qihao Zhu}, \bibinfo{person}{Zeyu Sun}, \bibinfo{person}{Yuan-an Xiao}, \bibinfo{person}{Wenjie Zhang}, \bibinfo{person}{Kang Yuan}, \bibinfo{person}{Yingfei Xiong}, {and} \bibinfo{person}{Lu Zhang}.} \bibinfo{year}{2021}\natexlab{}.
\newblock \showarticletitle{A syntax-guided edit decoder for neural program repair.}, In \bibinfo{booktitle}{ESEC/FSE '21: 29th ACM Joint European Software Engineering Conference and Symposium on the Foundations of Software Engineering, Athens, Greece, August 23-28, 2021}, \bibfield{editor}{\bibinfo{person}{Diomidis Spinellis}, \bibinfo{person}{Georgios Gousios}, \bibinfo{person}{Marsha Chechik}, {and} \bibinfo{person}{Massimiliano~Di Penta}} (Eds.).
\newblock \bibinfo{journal}{\emph{ESEC/SIGSOFT FSE}}, \bibinfo{pages}{341--353}.
\newblock
\urldef\tempurl%
\url{https://arxiv.org/pdf/2106.08253}
\showURL{%
\tempurl}


\end{thebibliography}
\end{multicols}

\end{document}